\documentclass[pdflatex,sn-mathphys-num]{sn-jnl}

\usepackage{graphicx}%
\usepackage{multirow}%
\usepackage{amsmath,amssymb,amsfonts}%
\usepackage{amsthm}%
\usepackage{mathrsfs}%
\usepackage[title]{appendix}%
\usepackage{xcolor}%
\usepackage{textcomp}%
\usepackage{subfigure}
\usepackage{manyfoot}%
\usepackage{booktabs}%
\usepackage{algorithm}%
\usepackage{algorithmicx}%
\usepackage{algpseudocode}%
\usepackage{listings}%
\usepackage{tikz}
\usepackage{bm}
\usepackage[T1]{fontenc}

\newcommand{\beq} {\begin{equation}}
\newcommand{\eeq} {\end{equation}}
\newcommand{\bdm} {\begin{displaymath}}
\newcommand{\edm} {\end{displaymath}}
\newcommand{\bit}{\begin{itemize}}
\newcommand{\eit}{\end{itemize}}
\newcommand{\bde}{\begin{description}}
\newcommand{\ede}{\end{description}}
\newcommand{\bce}{\begin{center}}
\newcommand{\ece}{\end{center}}
\newcommand{\ben} {\begin{enumerate}}
\newcommand{\een} {\end{enumerate}}
\newcommand{\bea} {\begin{eqnarray}}
\newcommand{\eea} {\end{eqnarray}}
\newcommand{\barr} {\begin{array}}
\newcommand{\earr} {\end{array}}
\newcommand{\bean} {\begin{eqnarray*}}
\newcommand{\eean} {\end{eqnarray*}}
\newcommand{\edoc} {\end{document}}

\newcommand{\eval}[2][\right]{\relax
  \ifx#1\right\relax \left.\fi#2#1\rvert}

\newcommand{\mc}[1]{\mathcal{#1}}

\newcommand{\mb}[1]{\bm{#1}}

\newcommand{\LRp}[1]{\left( #1 \right)}
\newcommand{\LRs}[1]{\left[ #1 \right]}

\newcommand{\LRc}[1]{\left\{ #1 \right\}}

\newcommand{\bigO}{\mathcal{O}}

\newcounter{tablecell}

\newcommand{\figlab}[1]{\label{fig:#1}}
\newcommand{\eqnlab}[1]{\label{eq:#1}}

\newcommand{\figref}[1]{\ref{fig:#1}}

\newcommand{\eqnref}[1]{\eqref{eq:#1}}

\newcommand{\seclab}[1]{\label{sect:#1}}
\newcommand{\secref}[1]{\ref{sect:#1}}

\newcommand{\rhok}{\rho_{\mb{k}}}
\renewcommand{\v}{v}

\newcommand{\kb}{\mb{k}}

\newcommand{\Sk}{S_{\mb{k}}}

\newcommand{\vb}{\mb{\v}}

\newcommand{\xb}{\mb{x}}

\newcommand{\B}{\mb{B}}

\newcommand{\Eb}{\mb{E}}

\newcommand{\Ebk}{\Eb_{\mb{k}}}

\renewcommand{\P}{P}

\theoremstyle{thmstyleone}%
\theoremstyle{thmstyletwo}%

\theoremstyle{thmstylethree}%

\algnewcommand{\P2P}{\textsc{P2P}}
\algnewcommand{\Alltoall}{\textsc{Alltoall}}
\algnewcommand{\Allreduce}{\textsc{Allreduce}}

\begin{document}

\title[Distributed Parallelization of PIF]{On Distributed Parallelization Strategies for Particle-in-Fourier Schemes}


\author*[1]{\fnm{Sriramkrishnan} \sur{Muralikrishnan}}\email{s.muralikrishnan@fz-juelich.de}

\author[2]{\fnm{Paul} \sur{Fischill}}\email{pfischill@ethz.ch}

\author[3,2]{\fnm{Andreas} \sur{Adelmann}}\email{andreas.adelmann@psi.ch}

\author[1]{\fnm{Robert} \sur{Speck}}\email{r.speck@fz-juelich.de}

\affil*[1]{\orgname{Jülich Supercomputing Centre}, \orgaddress{\city{Jülich}, \postcode{52428}, \country{Germany}}}

\affil[2]{\orgname{ETH Zürich}, \orgaddress{\city{Zürich}, \postcode{8092}, \country{Switzerland}}}

\affil[3]{\orgname{Paul Scherrer Institute}, \orgaddress{\city{Villigen}, \postcode{5232}, \country{Switzerland}}}


\abstract{We present and compare distributed parallelization strategies for the particle-in-Fourier (PIF) schemes used in kinetic plasma simulations. The different strategies are i) domain decomposition, where both the particles and Fourier modes are split between the MPI ranks ii) particle decomposition, where only the particles are split between the ranks and each rank carries all the modes, and, iii) space-time decomposition, in which time parallelization based on the parareal algorithm is added on top of the particle decomposition. We describe the different communication patterns involved in each of the strategies, the parameter regimes where they work best, and explain their advantages and disadvantages. We implement the strategies within the open-source, performance portable library IPPL and conduct scaling studies with 3D-3V Landau damping and Penning trap benchmark problems on Alps and JUWELS booster supercomputers. 
We analyze the dominant component timings in each of the strategies and identify areas for future optimizations.}


\maketitle

\section{Introduction}
\seclab{intro}

Kinetic plasma simulations play an important role in modeling nuclear fusion, particle accelerators and astrophysical phenomena among others. Particle-based numerical methods, specifically particle-in-cell (PIC) schemes, are often the method of choice for these simulations \cite{birdsall2004plasma,hockney2021computer}. However, the standard explicit PIC scheme is not energy conserving and suffers from aliasing, which leads to numerical instabilities such as the finite-grid instability and grid heating \cite{langdon1970effects,birdsall2004plasma,huang2016finite}. Mitigating these effects necessitates the use of a large number of grid points and particles, which in turn significantly increases the computational cost of these simulations. 

This urged the kinetic plasma physics community to study energy-conserving and/or structure-preserving PIC schemes, which either employ implicit time integrators \cite{markidis2011energy,chen2011energy} or are based on a variational formulation \cite{squire2012geometric,jianyuan2018structure,campos2024variational,campos2022variational} and discretization of the underlying Hamiltonian structure \cite{he2016hamiltonian,kraus2017gempic}. However, these schemes still do not fully eliminate aliasing and do not simultaneously conserve energy and momentum. They also differ significantly from the standard PIC schemes and hence may not be intuitive or easy to transition to from an application and implementation point of view.

Particle-in-Fourier (PIF) schemes are a promising alternative to PIC schemes, offering spectral accuracy with excellent long-time stability and conservation properties \cite{evstatiev2013variational,mitchell2019efficient}. They retain the general simplistic structure of standard PIC schemes and can hence be implemented in production PIC codes in a relatively straightforward manner. While originally limited to periodic boundary conditions, they have recently been extended to free space and Dirichlet boundary conditions in \cite{shen2024particle}.

Parallelization strategies for PIF schemes are less studied than those for PIC schemes, where extensive literature \cite{walker1990characterizing,carmona1997parallel,di2001parallel} and numerous open-source codes \cite{vay2018warp,burau2010picongpu,muralikrishnan2024scaling} are available. Shared memory, CPU-based parallelism for PIF schemes is considered in \cite{mitchell2019efficient}. In terms of distributed parallelization strategies, particle decomposition, where only the particles are divided between ranks and the modes are duplicated in each rank, is considered in \cite{briguglio2000parallelization,muralikrishnan2025parapif}. Domain decomposition, a common distributed parallelization strategy for PIC schemes, where both modes and particles are divided between ranks, needs distributed nonuniform fast Fourier transforms (NUFFTs) when applied to PIF schemes. This has been pursued in the context of CPU-based molecular dynamics simulations in \cite{pippig2013parallel,pippig2016massively}. Recently, time parallelization and space-time parallelization, where time parallelization is added on top of particle decomposition, have been introduced in \cite{muralikrishnan2025parapif}. 

These parallelization strategies differ in i) their level of intrusiveness for implementation in production-grade particle-mesh codes, ii) their communication patterns and efficiencies, and iii) the parameter regimes (modes, particles and number of time steps) where each of them work the best. In regimes where all of them are applicable, it is not clear which strategy offers the best scalability and time to solution. The goal of this paper is to fill this gap in the PIF literature by addressing the above questions. This will help researchers to pick the best strategy corresponding to their problem at hand and will also serve as a baseline upon which innovations and optimizations in each of the strategy can be developed.

The outline of this paper is as follows. We introduce the PIF scheme in Section 2. The different distributed parallelization strategies are introduced in Section 3. A comparison of the strategies in terms of their best parameter regimes, communication patterns and implementation aspects in a production particle-mesh code is shown in Section 4. The strong scaling results for a representative problem size comparing the different parallelization strategies for Landau damping and Penning trap benchmark problems are shown in Section 5. Finally, we present the conclusions and possible future works in Section 6.

\section{Particle-in-Fourier schemes}
\seclab{pif}

We consider the non-relativistic Vlasov-Poisson system in kinetic plasma simulations with a fixed magnetic field and introduce the PIF method in that setting. 

The electrons are immersed in a uniform, immobile, neutralizing background ion population, and the electron dynamics is given by
    \begin{equation}
        \eqnlab{Vlasov}
        \frac{\partial f}{\partial t} + \vb \cdot \nabla_{\xb} f + \frac{q_e}{m_e}\LRp{\Eb + \vb \times \B_{ext}} \cdot \nabla_{\vb} f = 0,
    \end{equation}
where $\Eb = \Eb_{sc} + \Eb_{ext}$ is the total electric field, and $\Eb_{ext}$ and $\B_{ext}$ are the known external electric and magnetic fields. We use boldfaced letters for vector-valued quantities and non-bold ones for scalar quantities. Here, $f(\xb,\vb,t)$ is the electron phase-space distribution function and $q_e$, $m_e$ are the electron charge and mass respectively. The total
electron charge in the system is given by $Q_e=q_e\int\int f d\xb d\vb$, the electron charge density by $\rho_e(\xb) = q_e\int f d\vb$ and the
constant ion density by $\rho_i = Q_e/\int d\xb$. Let us denote the permittivity of free space by $\varepsilon_0$. The self-consistent potential ($\phi$) and the electric field ($\Eb_{sc}$) due to space charge are given by 
\begin{align}
    \eqnlab{potential}
    \quad -\Delta \phi &= \rho/\varepsilon_0 = \LRp{\rho_e - \rho_i}/\varepsilon_0, \\
    \Eb_{sc} &= -\nabla \phi.
\end{align}

The PIF method discretizes the phase-space distribution $f(\xb,\vb,t)$ in a Lagrangian manner using macro-particles (hereafter referred to as ``particles'' for simplicity) as 
\begin{equation}
 \eqnlab{dist_f}
     f\LRp{\xb,\vb,t} = \sum_{j=1}^{N_p} w_j S\LRp{\xb-\xb_j} \delta\LRp{\vb-\vb_j},
 \end{equation}
 where $w_j$ is the particle weight, $S\LRp{\xb-\xb_j}$ is the shape function in configuration space and $\delta\LRp{\vb-\vb_j}$ is the Dirac-delta distribution in velocity space. At time $t=0$, the
distribution $f$ is sampled, which leads to the creation of particles. Subsequently, a typical computational cycle in PIF consists of the following steps:

\begin{enumerate}
    \item \label{step1_pif} Assign a shape function --- e.g., cloud-in-cell \cite{birdsall2004plasma} --- to each particle $j$ and deposit the electron charge onto the Fourier space which gives 
    \begin{align} 
\rhok\LRp{\kb} &= \frac{1}{L^3}\int\rho\LRp{\xb}\exp\LRp{-i\kb\cdot\xb} d\xb \\
\eqnlab{scatter_pif}
            &\approx \frac{q_e\Sk}{L^3}\sum_{j=1}^{N_p}\exp\LRp{{-i\kb\cdot\xb_j}}.
\end{align}
    Here, for simplicity of exposition, we consider three dimensions. We denote by $N$ the number of Fourier modes in each dimension, by $N_m = N^3$ the total number 
of modes and by $L$ the the length of the domain in each dimension. We 
also assume $N$ to be even and define the Fourier modes $\kb \in K_N = \LRc{\frac{2\pi}{L}[-N/2,\ldots,0,\ldots,N/2-1]}^3$. $\Sk=\mc{F}\LRp{S(x)}$ is the Fourier transform of the shape function $S$, which is usually available in analytic form.
    \item \label{step2_pif} The Poisson equation is solved using the charge density in Fourier space, and the electric field is obtained by $\Ebk = -\frac{i\kb}{|\kb|^2} \rhok$.
    \item \label{step3_pif} The electric field is gathered from Fourier space to the particle positions using
 \begin{equation}
\eqnlab{gather_pif}
\Eb(\xb_j) = \sum_{\kb\in K_N}\Ebk\Sk\exp\LRp{{i\kb\cdot\xb_j}}.
\end{equation}
    \item \label{step4_pif} By means of a time integrator, advance the particle positions and velocities using
        \begin{equation}
        \eqnlab{velposPIF}
            \frac{d\vb_j}{dt} = \frac{q_e}{m_e}\LRp{\Eb + \vb \times \B_{ext}}|_{\xb=\xb_j}, \quad
            \frac{d\xb_j}{dt} = \vb_j.
        \end{equation}
\end{enumerate}

A Naive calculation of the interpolation from particles to Fourier modes and vice versa in steps \ref{step1_pif} and \ref{step3_pif} has a computational complexity $\bigO\LRp{N_pN_m}$, which is prohibitively expensive except when only a small number of particles and Fourier modes are considered. Hence, the key step introduced in \cite{mitchell2019efficient} that makes PIF schemes practical is the use of NUFFTs of type 1 and 2 \cite{dutt1993fast,dutt1995fast,potts2001fast,barnett2019parallel} to compute equations
\eqnref{scatter_pif} and \eqnref{gather_pif}, respectively. This reduces the complexity of PIF schemes to $\mc{O}\LRp{\LRp{|log\varepsilon|+1}^dN_p+N_mlogN_m}$, which is comparable to the commonly used PIC schemes although with a larger constant in front of $N_p$. The constant depends on the tolerance $\varepsilon$ chosen for the NUFFT, as---unlike the uniform FFT---the NUFFT is an approximate algorithm. 

PIF differs from the standard FFT PIC algorithm in steps 1 and 3, where it interpolates directly from the particles to Fourier space and vice versa, avoiding a grid in real space. The complexity of the standard FFT PIC algorithm with the commonly used linear B-splines for interpolations is $\mc{O}\LRp{2^dN_p+N_mlogN_m}$. Thus, PIF can be considered as a fully spectral version of the commonly used lower order PIC algorithm.  

\section{Parallelization Strategies for PIF schemes}
\seclab{different_parallel_pif}

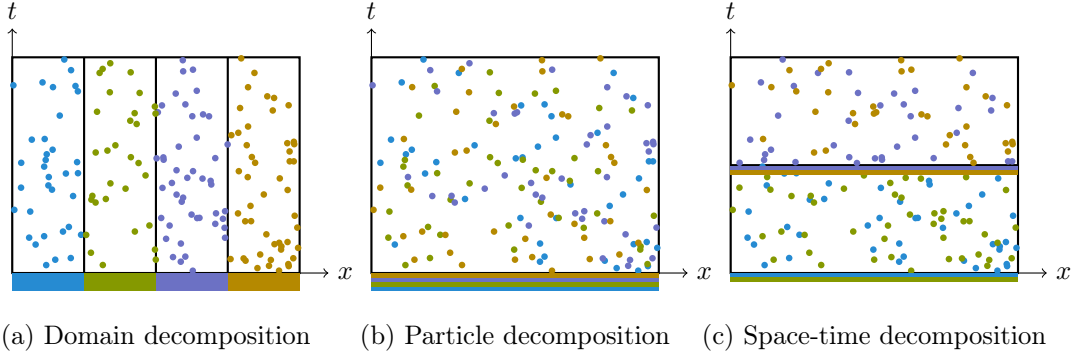
\begin{figure}[h!b!t!]
\centering
\begin{tikzpicture}[scale=0.95]

\def\W{4}      
\def\H{3}      
\def\barh{0.25}
\def\dx{5}

\definecolor{r0}{RGB}{38,139,210}   
\definecolor{r1}{RGB}{133,153,0}    
\definecolor{r2}{RGB}{108,113,196}  
\definecolor{r3}{RGB}{181,137,0}    

\def\Np{160} 

\pgfmathsetseed{12345} 

\foreach \i in {1,...,\Np} {
    \pgfmathsetmacro{\tmpx}{rnd*\W}
    \pgfmathsetmacro{\tmpt}{rnd*\H}
    \expandafter\xdef\csname px\i\endcsname{\tmpx}
    \expandafter\xdef\csname pt\i\endcsname{\tmpt}
}

\newcommand{\drawparticles}[3]{%
  \foreach \i in {#1,...,#2} {
    \fill[#3]
      (\csname px\i\endcsname, \csname pt\i\endcsname)
      circle (1.3pt);
  }
}

\begin{scope}

\draw[->] (0,0) -- (\W+0.4,0) node[right] {$x$};
\draw[->] (0,0) -- (0,\H+0.4) node[above] {$t$};

\draw[thick] (0,0) rectangle (\W,\H);

\foreach \i in {1,2,3}
  \draw[thick] (\i*\W/4,0) -- (\i*\W/4,\H);

\foreach \i in {1,...,\Np} {
  \pgfmathparse{\csname px\i\endcsname}
  \let\x=\pgfmathresult

  \ifdim \x pt < 1pt
    \def\col{r0}
  \else\ifdim \x pt < 2pt
    \def\col{r1}
  \else\ifdim \x pt < 3pt
    \def\col{r2}
  \else
    \def\col{r3}
  \fi\fi\fi

  \fill[\col] (\x, \csname pt\i\endcsname) circle (1.3pt);
}

\fill[r0] (0,-\barh) rectangle (1,0);
\fill[r1] (1,-\barh) rectangle (2,0);
\fill[r2] (2,-\barh) rectangle (3,0);
\fill[r3] (3,-\barh) rectangle (4,0);

\node at (\W/2,-0.9) {(a) Domain decomposition};

\end{scope}

\begin{scope}[xshift=\dx cm]

\draw[->] (0,0) -- (\W+0.4,0) node[right] {$x$};
\draw[->] (0,0) -- (0,\H+0.4) node[above] {$t$};

\draw[thick] (0,0) rectangle (\W,\H);

\drawparticles{1}{40}{r0}
\drawparticles{41}{80}{r1}
\drawparticles{81}{120}{r2}
\drawparticles{121}{160}{r3}

\fill[r0] (0,-\barh) rectangle (\W,-0.75*\barh);
\fill[r1] (0,-0.75*\barh) rectangle (\W,-0.5*\barh);
\fill[r2] (0,-0.5*\barh) rectangle (\W,-0.25*\barh);
\fill[r3] (0,-0.25*\barh) rectangle (\W,0);

\node at (\W/2,-0.9) {(b) Particle decomposition};

\end{scope}

\begin{scope}[xshift=2*\dx cm]

\draw[->] (0,0) -- (\W+0.4,0) node[right] {$x$};
\draw[->] (0,0) -- (0,\H+0.4) node[above] {$t$};

\draw[thick] (0,0) rectangle (\W,\H);

\draw[thick] (0,\H/2) -- (\W,\H/2);


\foreach \i in {1,...,\Np} {

  \pgfmathparse{\csname pt\i\endcsname}
  \let\t=\pgfmathresult

  \ifdim \t pt < 1.5pt
    \ifnum \i<81
      \def\col{r0}
    \else
      \def\col{r1}
    \fi
  \else
    \ifnum \i<81
      \def\col{r2}
    \else
      \def\col{r3}
    \fi
  \fi

  \fill[\col] (\csname px\i\endcsname, \t) circle (1.3pt);
}

\fill[r0] (0,-0.25*\barh) rectangle (\W,0);
\fill[r1] (0,-0.5*\barh) rectangle (\W,-0.25*\barh);
\fill[r2] (0,1.485-0.25*\barh) rectangle (\W,1.485);
\fill[r3] (0,1.485-0.5*\barh) rectangle (\W,1.485-0.25*\barh);

\node at (\W/2,-0.9) {(c) Space-time decomposition};

\end{scope}
\end{tikzpicture}
\caption{Schematic of the three parallelization strategies for particle-in-Fourier schemes.
Colors denote space-time MPI ranks, which is $4\times1$ for domain and particle decompositions and $2\times2$ for space-time decomposition. Horizontal cuts indicate time decomposition; vertical cuts indicate spatial decomposition. The colored bars denote Fourier modes which are distributed for domain decomposition and duplicated for particle and space-time decompositions.}
\figlab{illustration}
\end{figure}

\subsection{Domain decomposition}

A common strategy for spatially parallelizing standard particle-mesh methods, such as PIC schemes, is domain 
decomposition. Although formally PIF schemes avoid creating a grid in real space, the type 1 and type 2 NUFFTs used in them perform spreading and interpolation operations to/from an upsampled grid using higher order kernels. Hence, we can apply domain decomposition to distribute them, similiarly to PIC schemes. In this approach, the spatial domain is divided between
multiple MPI ranks\footnote{Throughout this discussion, whenever we refer to MPI ranks, the discussion applies equally to CPU nodes as well and accelerators such as GPUs, as we consider a one-to-one mapping between them. In the case of CPU nodes, in addition to distributed parallelism we use shared memory parallelism within the node.} and each rank holds a small portion of that domain. It also stores the particles which are 
spatially contained within that local domain. Each local
domain also contains a certain number of halo or ghost layers, depending
on the width of the shape functions used for the spreading and
interpolation operations. The particles within that local domain
scatter to it, and the halo layers are then exchanged with
neighboring MPI ranks to add contributions from them. The Poisson equation is solved in a distributed manner. After computing the electric field, the halo layers are exchanged once again to get the correct field quantities from neighboring subdomains. Finally, 
interpolation from the grid to the particles is performed locally using
data from the domain of the rank along with its halo layers. Once the
particle positions and velocities are updated, the particles
leaving the current subdomain are sent to the MPI rank owning the corresponding neighbouring subdomain. The field and
particle operations are fully distributed in this parallelization, and it works well for uniform and mildly nonuniform particle
distributions. However, for highly nonuniform and clustered particle distributions, particle load balancing strategies are crucial to 
distribute the particles more or less equally to the processors. These particle load
balancing strategies typically lead to load imbalance in the fields and adversely affect the strong and weak scaling \cite{muralikrishnan2024scaling}. 
Hence, load balancing remains a very active research topic in domain decomposition \cite{muralikrishnan2024scaling,rowan2021situ}. A pseudo-code for the PIF scheme with domain decomposition is given in Algorithm \ref{alg:domain-decomp}.

\begin{algorithm}[t]
\caption{Particle-in-Fourier with domain decomposition. Point-to-point (P2P) communication is involved for the particle migration as well as halo communication for the fields. In addition, the distributed FFTs in NUFFTs involve Alltoall communication on fields.}
\label{alg:domain-decomp}
\begin{algorithmic}[1]


\State Distribute particles spatially across ranks
\State Distribute Fourier modes across ranks

\For{$n = 0$ to $N_t - 1$}

    \State \textbf{Particle push:}
    \State \quad Advance $(\xb_j, \vb_j)$

    \State \textbf{Particle migration:}
    \State \quad Exchange particles crossing subdomain boundaries \Comment{P2P} particles

    \State \textbf{Charge deposition:}
    \State \quad Perform charge deposition with type-1 NUFFT \Comment{P2P}, Alltoall density field

    \State \textbf{Field solve:}
    \State \quad Solve Poisson equation in Fourier space

    \State \textbf{Field interpolation:}
    \State \quad Interpolate electric field with type-2 NUFFT \Comment{P2P}, Alltoall electric field

\EndFor
\end{algorithmic}
\end{algorithm}

\subsection{Particle decomposition}
In a domain-decomposed implementation, NUFFT-based PIF typically requires wider
spreading/interpolation stencils than standard PIC that uses lower order B-spline shape
functions. This follows from the fact that the support size of optimal NUFFT
window functions scale as $w=\mathcal{O}(\log(1/\varepsilon))$. For instance, with $w \approx \lvert \log_{10}\varepsilon\rvert + 1$, a tolerance of $\varepsilon=10^{-7}$
corresponds to $w=8$ grid points, while $\varepsilon=10^{-16}$ corresponds to $w=17$.
By contrast, commonly used lower order B-splines in PIC typically require only $1$--$2$
halo layers for charge deposition and field interpolation. Consequently, the halo
exchanges associated with the NUFFT spreading/interpolation can be noticeably more
expensive than in standard PIC.

This motivates an alternative distribution strategy for PIF: \emph{pure particle
decomposition}~\cite{briguglio2000parallelization,mitchell2019efficient}. Here, the
particles are partitioned across MPI ranks, while each rank holds all modes of the
Fourier-space fields. After computing the Fourier-space charge density
from the local particles (scatter/type-1 NUFFT), the ranks perform a global sum
reduction (e.g., \texttt{MPI\_Allreduce}) to obtain the total $\rho_k$. The subsequent
steps---the field solve in Fourier space (step~2) and the gather/type-2 NUFFT back to
particle locations (step~3)---can then be carried out locally, since the Fourier data is
replicated on every rank. A key advantage of this approach is that it is load balanced
by construction for both particles and fields, and it remains robust under highly
nonuniform or clustered particle distributions, because particles do not migrate between
spatial subdomains.

Particle decomposition is attractive only when the total number of Fourier modes is
small enough that replicating them is affordable. The dominant cost of PIF is the NUFFT,
with complexity
$\bigO\!\left((\lvert\log\varepsilon\rvert+1)^d N_p + N_m\log N_m\right)$.
Under particle decomposition, the $N_m\log N_m$ work and the storage associated with
the modes do not decrease with the number of MPI ranks, creating a serial
bottleneck as the simulation is strong-scaled. In addition, the \texttt{MPI\_Allreduce} over the
Fourier modes after the scatter is a global collective, whose cost increases with both the
number of modes and the number of ranks further limiting the scalability. A pseudo-code for the PIF scheme with particle decomposition is given in Algorithm \ref{alg:particle-decomp}.

\begin{algorithm}[t]
\caption{Particle-in-Fourier with particle decomposition. Only Allreduce communication is needed for the density field whereas all other operations happen locally due to the duplication of the fields.}
\label{alg:particle-decomp}
\begin{algorithmic}[1]


\State Distribute particles spatially across ranks
\State Duplicate Fourier modes across ranks

\For{$n = 0$ to $N_t - 1$}

    \State \textbf{Particle push:}
    \State \quad Advance $(\xb_j, \vb_j)$

    \State \textbf{Charge deposition:}
    \State \quad Perform charge deposition locally with type-1 NUFFT
    \State \quad All-reduce charge density across ranks \Comment{Allreduce} density field

    \State \textbf{Field solve:}
    \State \quad Solve Poisson equation in Fourier space

    \State \textbf{Field interpolation:}
    \State \quad Interpolate electric field locally with type-2 NUFFT 

\EndFor
\end{algorithmic}
\end{algorithm}

\subsection{Space-time decomposition}

When scaling through spatial parallelization saturates, further reductions in
time to solution and scalability of the simulation can be achieved by exploiting parallelism in the time integration. Parallel-in-time (PinT) 
algorithms have a long history, starting from the pioneering work of 
Nievergelt in 1964 \cite{nievergelt1964parallel}. A comprehensive review of the different
types of PinT algorithms can be found in \cite{gander201550,ong2020applications,pintwebsite}. The simplest and most widely investigated PinT algorithm, parareal \cite{lions2001resolution}, is an iterative scheme that can be thought of as a predictor-corrector approach. The initial time domain is split into multiple intervals and a cheap (possibly inaccurate) predictor called ``coarse propagator'' ($G$) is run serially to give initial guesses for an accurate and expensive corrector called ``fine propagator'' ($F$). With these
guesses, the fine propagator is run in parallel in each of these intervals. The parareal correction step is performed at the end of each iteration as follows:
\begin{align}
    U_0^{k+1} &:= u ^0, \\
\eqnlab{parareal_correction}
    U_{n+1}^{k+1} &:=
    F(T_{n+1}, T_n, U_n^k)
    + G(T_{n+1}, T_n, U_n^{k+1})
    - G(T_{n+1}, T_n, U_n^{k}),
\end{align}
where, $[T_n,T_{n+1}]$ are the time intervals, $\Delta T$ is the size of the time subdomain, $U_n^{k}$ is the approximation of the solution at $T_n$ in the $k^{\text{th}}$ parareal iteration, and $u^0$ is the initial condition. The iterations are performed until convergence to a specific tolerance.

A novel parareal based time parallelization strategy for PIF schemes has been recently developed in \cite{muralikrishnan2025parapif}. In this scheme, given a PIF scheme with NUFFT of tolerance $\varepsilon_f$ as a fine propagator, we use either the PIF scheme with a coarse NUFFT tolerance $\varepsilon_g > \varepsilon_f$ or the standard PIC scheme as the coarse propagator in the parareal algorithm. Both PIF and PIC coarse propagators may or may not employ time coarsening depending on the fine time step size and the stability of the time integrators. The convergence of the parareal algorithm for PIF with both the coarse propagators is shown in \cite{muralikrishnan2025parapif} and the choice depends on the specific application. More details and a heuristic guidance on selecting the suitable coarse propagator can be found in \cite{muralikrishnan2025parapif}. This time parallelization approach is added on top of the particle decomposition spatial parallelization to create a space-time decomposition in \cite{muralikrishnan2025parapif}. Particle decomposition is chosen here because the algorithm requires tracking the same set of particles in each MPI rank during different time intervals, which is not easily achievable using the domain decomposition approach. In domain decomposition, the same particle in different time slices can be in different ranks. Hence, adding time parallelization on top of this is non-trivial and involves communication across different space and time ranks. This approach has not been pursued yet and deserves a separate study in future work. A pseudo-code for the PIF scheme with space-time decomposition is given in Algorithm \ref{alg:spacetime-decomp}. For brevity, we do not show the multiblock parareal approach in Algorithm \ref{alg:spacetime-decomp} and refer the readers to \cite{muralikrishnan2025parapif,aubanel2011scheduling,nielsen2018communication} for more details on this. In Figure \figref{illustration}, we show a schematic comparing the three parallelization strategies for the same particle evolution. 

\begin{algorithm}[t]
\caption{Particle-in-Fourier with space-time decomposition. In addition to the Allreduce communication on the density field similar to particle decomposition, there is also P2P communication of particles across the time slabs owing to time parallelization. The fine propagator is PIF with particle decomposition whereas the coarse propagator is PIF or PIC with particle decomposition.}
\label{alg:spacetime-decomp}
\begin{algorithmic}[1]

\State Initialize MPI ($P$ = $P_s$ × $P_t$ ranks)
\State Distribute particles across $P_s$ MPI ranks
\State Split time domain into $P_t$ slabs
\State Duplicate Fourier modes across all MPI ranks
\State Run coarse propagator sequentially to get initial particle states in all $P$ ranks 
\While{not converged}
    \State Run fine propagator in parallel to get $F(U_n^k)$ \Comment{Allreduce} density field
    \State Perform $D(U_n^k) = F(U_n^k)-G(U_n^k)$ locally from the saved $G(U_n^k)$
    \If{rankTime > 0}
    \State Receive $U_n^{k+1}$ from rankTime-1 \Comment{P2P} particles in time
    \EndIf
    \State Run coarse propagator to get $G(U_n^{k+1})$ \Comment{Allreduce} density field
    \State Perform parareal correction $U_{n+1}^{k+1} = G(U_n^{k+1}) + D(U_n^k)$
    \State Compute convergence criterion
    \If{rankTime < $P_t-1$}
    \State Send $U_n^{k+1}$ to rankTime \Comment{P2P} particles in time
    \EndIf
\EndWhile
\end{algorithmic}
\end{algorithm}

\section{Comparison of parallelization strategies}


\subsection{Suitable parameter regimes}

Domain decomposition works best when both the number of Fourier modes and the total number of particles required in an application are high. In fact, it is the only strategy for which the total number of Fourier modes is not limited by the memory available in a single rank. Domain decomposition needs to be accompanied with suitable particle and field load balancing strategies for handling nonhomogeneous distributions. Achieving load balance in both particles and fields is a nontrivial task and, at the time of writing, is an active area of research.

Particle decomposition is the best parallelization strategy when the number of Fourier modes needed is small in relation to the total number of particles. This is because communication happens only via Fourier modes in particle decomposition and hence it avoids communicating a large volume of particles between ranks. It is also load balanced in both particles and fields by design and can handle nonhomogeneous distributions in the same way as homogeneous distributions. However, when a large number of Fourier modes are needed, it is limited by the memory available in one rank/GPU and hence it is not possible in those cases or not preferable due to high serial computational costs. 

Space-time decomposition, in which time parallelization is added on top of particle decomposition, can help to get additional speedup compared to spatial parallelization alone. It works the best when a lot of time steps are needed in a simulation and the spatial parallelization is either not effective (small problem size) or has saturated. However, the effectiveness of the parareal method varies between test cases. It should also be noted that in some applications, like the Penning trap shown in Section 5, time parallelization works effectively only within a small time window. Hence, in those cases, a large number of time steps does not directly translate to potential for large speedups with time parallelization. Another important difference in iterative time parallelization strategies like parareal compared to spatial parallelization strategies is that the results match with the serial time stepping results only up to the tolerance set for the convergence of the parareal method. Choosing a proper tolerance in time parallelization is often not a straightforward task and, if not done carefully, can lead to exaggerated speedups compared to serial time stepping as shown in \cite{gotschel2020twelve}.

\subsection{Communication patterns}
\begin{table}[t]
\centering
\caption{Communication patterns for different parallelization strategies.
``Collective'' refers to global reductions or Alltoall operations, while
``P2P'' denotes point-to-point communication between neighboring ranks.}
\label{tab:comm-patterns}

\begin{tabular}{l|cc|cc}
\hline
 & \multicolumn{2}{c|}{\textbf{Particles}} 
 & \multicolumn{2}{c}{\textbf{Fields (Fourier modes)}} \\
\textbf{Strategy}
 & Space & Time & Space & Time \\
\hline
Domain decomposition
 & P2P
 & none 
 & P2P, Collective (Alltoall)
 & none \\
Particle decomposition
 & none
 & none
 & Collective (Allreduce)
 & none \\
Space--time decomposition
 & none
 & P2P
 & Collective (Allreduce)
 & none \\
\hline
\end{tabular}
\end{table}

In Table \ref{tab:comm-patterns}, we list the communication patterns involved in space and time for particles and fields in all the strategies. Domain decomposition involves point-to-point communication between spatial neighboring ranks for both particles and halo cell exchanges in fields communication. In addition, the uniform FFTs in distributed NUFFTs involve collective (Alltoall) communication in the fields. Since both particles and modes are distributed it is work optimal and memory optimal up to halo layers. However, it involves a dynamic communication pattern, as the volume of particles communicated to the neighboring ranks changes depending on the phase-space evolution in the problem. This poses challenges in buffer pre-allocation and management, which is critical for scalability and performance on GPUs.

Particle decomposition involves the simplest communication pattern among the three strategies and requires only a single collective operation (Allreduce) on the Fourier modes corresponding to the charge density. Otherwise, all the other operations involving fields and particles are local, and hence do not need any communication. The communication pattern is thus static in terms of volume and does not need any explicit buffer management. However, it is not memory- and work-optimal for the field operations due to the replicated Fourier modes in each rank. 

Space-time decomposition involves point-to-point communication for the particles with the forward looking neighboring rank in the time communicator and collective (Allreduce) communication on the Fourier modes within the spatial communicator. As in particle decomposition, the communication pattern is fixed in terms of communication volume and neighbors. Hence, the buffer management is simple and all the buffers needed for point-to-point communications can be pre-allocated once without the need for resizing. However, it is also not work- and memory-optimal. In addition to the spatial redundancy from particle decomposition, the parareal method adds redundancy in computation and memory for the time parallelization. It is interesting to note that the particle and space-time decompositions convert the PIF method, which is Lagrangian by nature and involves dynamic communication patterns, to a simplified static communication pattern at the expense of redundant memory storage and computations.  

\subsection{Implementation in a particle-mesh code}

In this section, we discuss the implementation aspects of the PIF scheme, using the parallelization strategies discussed in Section \secref{different_parallel_pif}, within a production particle-in-cell code. Our discussion is based on an implementation in the open-source, performance portable C++ library IPPL \cite{muralikrishnan2024scaling,matthias_frey_2024_10878166}.

PIF differs from standard FFT PIC codes in the scatter and gather steps, which are replaced by NUFFTs. Particle decomposition is the simplest and least intrusive strategy in that respect as it needs only shared memory NUFFTs rather than distributed ones. There are many open-source NUFFT libraries that can be used for this purpose, such as FINUFFT \cite{barnett2019parallel,shih2021cufinufft} and NFFT3 \cite{keiner2009using}. In our studies, in addition to the native NUFFT implementation in IPPL which we use for domain decomposition, we have interfaced IPPL with FINUFFT which provides NUFFTs on CPUs and NVIDIA GPUs. Apart from the local NUFFTs, particle decomposition only needs a single \texttt{MPI\_Allreduce} call in the scatter step to convert the code to use distributed parallelism. 

Domain decomposition, on the other hand, requires distributed NUFFTs. Apart from our native NUFFT implementation in IPPL \cite{fischillnufft}, we are not aware of any other GPU-capable distributed NUFFT.
Hence, one may either use IPPL or implement NUFFT natively in their code base. The scatter and gather operations in type 1 and type 2 NUFFTs resemble closely the scatter and gather steps in PIC codes, albeit with a higher order shape function. Hence, if the codes already support higher order shape functions, most of the infrastructure required for distributed NUFFTs should already be in place. However, it still remains the most intrusive strategy among the three.

Space-time decomposition performs time parallelization on top of particle decomposition. The parareal method, which is what we use for time parallelization, is a non-intrusive time parallel strategy. The coarse propagators in the parareal algorithm are either PIF with coarser NUFFT tolerance or the standard PIC method, both of which should be readily available in production PIC codes. Thus, this strategy can be either directly implemented in the code base or realized using open-source libraries for time parallelization like XBraid \cite{falgout2014parallel}. Our results in Section 5 are based on the implementation directly in IPPL. Each field in IPPL can have their own MPI communicator, and this functionality has been used to write a unified code with minor branches for the implementation of all three parallelization strategies. 

\section{Scaling study on benchmark problems}

\subsection{Simulation setup}
We perform a strong scaling study and compare the parallelization strategies on 3D-3V weak Landau damping and Penning trap benchmark problems in kinetic plasma simulations. Landau damping has a nearly homogeneous particle distribution, whereas the particles are clustered in the Penning trap problem. The phase-space evolution of the particles in both test cases are representative of regimes happening in particle accelerator simulations. We choose $256^3$ Fourier modes and $10$ particles per mode, which results in roughly $168$ million total particles. We use a time step size of $\Delta t=0.003125$ and run for $6144$ time steps in case of space-time decomposition. For domain decomposition and particle decomposition, in order to decrease the necessary computing resources, we run for 768 time steps and scale it by a factor which accounts for the initialization costs. Generally, the run time of domain decomposition and particle decomposition scale linearly with the number of time steps with a constant of 1. However, due to initialization costs in the beginning of the simulation it can be less than one and depends on the specific machine in which we are running the simulation. This factor is obtained by running a few of the simulations for the whole 6144 time steps as well as for 768 time steps. Based on these simulations for domain decomposition, we use 6.5 (0.8125$\times$ ratio of time steps (6144/768) which is 8) and 6.8 (constant=0.85) as scaling factors on Alps and JUWELS booster respectively, whereas for particle decomposition we use a factor of 8 (constant=1) on both the systems. Finally, in the figures we report the wall time per simulation time step. The NUFFT tolerance for the PIF scheme is taken as $\varepsilon=10^{-7}$ and the stopping tolerance for the parareal algorithm in the space-time decomposition is taken as $10^{-8}$. These values are selected based on the study in \cite{muralikrishnan2025parapif} and related to the relative errors in the conservation of energy, momentum and charge for the selected time step size. 

The performance portable implementation is carried out in IPPL, and computations are performed on NVIDIA A100 and GH200 GPUs in the JUWELS Booster supercomputer at the Jülich Supercomputing Centre, Germany and on the daint partition of Alps supercomputer at CSCS, Switzerland, respectively. For particle decomposition, the density continuously decreases with increasing ranks and the current native NUFFT implementation in IPPL is not optimal for that setting. Hence, for particle and space-time decompositions we use FINUFFT \cite{shih2021cufinufft} interfaced with IPPL whereas for domain decomposition we use the native NUFFT implementation in it. We choose the output-driven option for type 1 NUFFT in both FINUFFT as well as the native NUFFT in IPPL, whereas for the type 2 NUFFT we use the sorted gather option in native NUFFT and the default option in FINUFFT. These are based on our numerical experiments with the two test cases. The parameters for the test cases are taken from \cite{muralikrishnan2025parapif}, and we briefly describe them here for the sake of completeness. 

\subsubsection{Landau damping}
Landau Damping is one of the classical benchmark problems in plasma physics. We consider the following initial distribution:
   \begin{equation}
        f(t=0) = \frac{1}{\LRp{2\pi}^{3/2}} e^{-|\vb|^2/2} \LRp{1+\alpha\cos(k x)} \LRp{1+\alpha\cos(k y)} 
               \LRp{1+\alpha\cos(k z)},
    \end{equation}
in the domain $\LRs{0,L}^3$, where $L = 2\pi/k$ is the length in each dimension. We choose $k = 0.5$, $\alpha=0.05$, which corresponds to weak Landau damping. The total electron charge based 
on our initial distribution is $Q_e = -L^3$.

\subsubsection{Penning trap}
This benchmark problem corresponds to the dynamics of electrons in a Penning trap with a neutralizing static ion background. Unlike the Landau damping test case, this one involves external electric and magnetic fields. The external magnetic field is given by $\B_{ext}=\LRc{0,0,5}$ and the quadrupole external electric field by 
    \begin{equation}
        \eqnlab{penning_ext_efield}
        \Eb_{ext} = \LRp{-\frac{15}{L}\LRp{x-\frac{L}{2}},-\frac{15}{L}\LRp{y-\frac{L}{2}},\frac{30}{L}\LRp{z-\frac{L}{2}}},
    \end{equation}
where the domain is $\LRs{0,L}^3$ and $L=25$. For the initial conditions, we sample the phase-space using a Gaussian
    distribution in all the variables. The mean and standard deviation for
    all the velocity components are $0$ and $1$, respectively. While the mean
    for all the configuration space variables is $L/2$, the standard 
    deviations are $2$, $1$ and $3$ for $x$, $y$, and $z$, respectively. The total electron charge is 
    $Q_e=-1562.5$.

\subsection{Total timings}

\begin{figure}[h!b!t!]
\subfigure[Time/time step Alps]{\includegraphics[width=0.5\columnwidth]{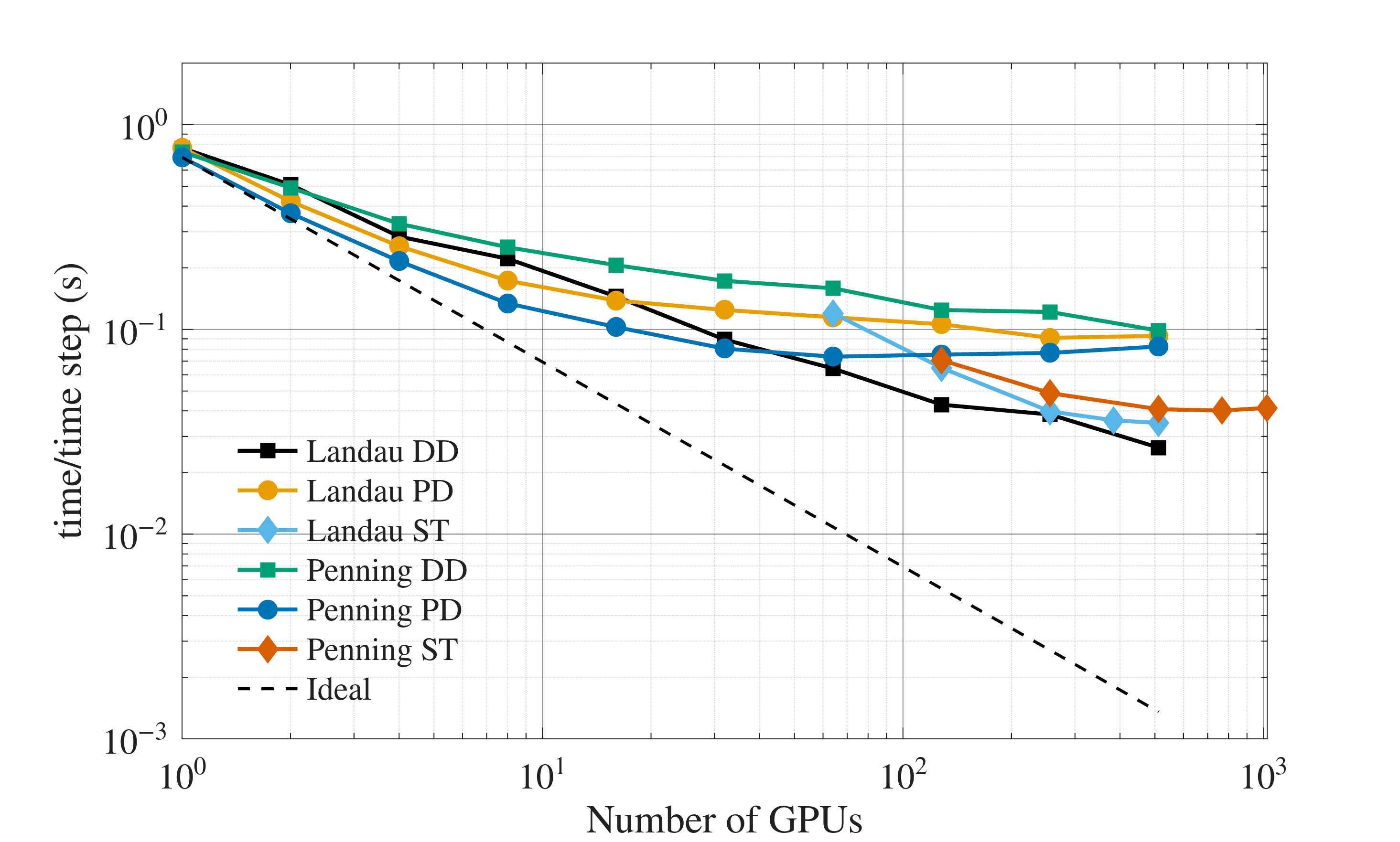}
\figlab{time_alps}}
\subfigure[Time/time step JUWELS Booster]{\includegraphics[width=0.5\columnwidth]{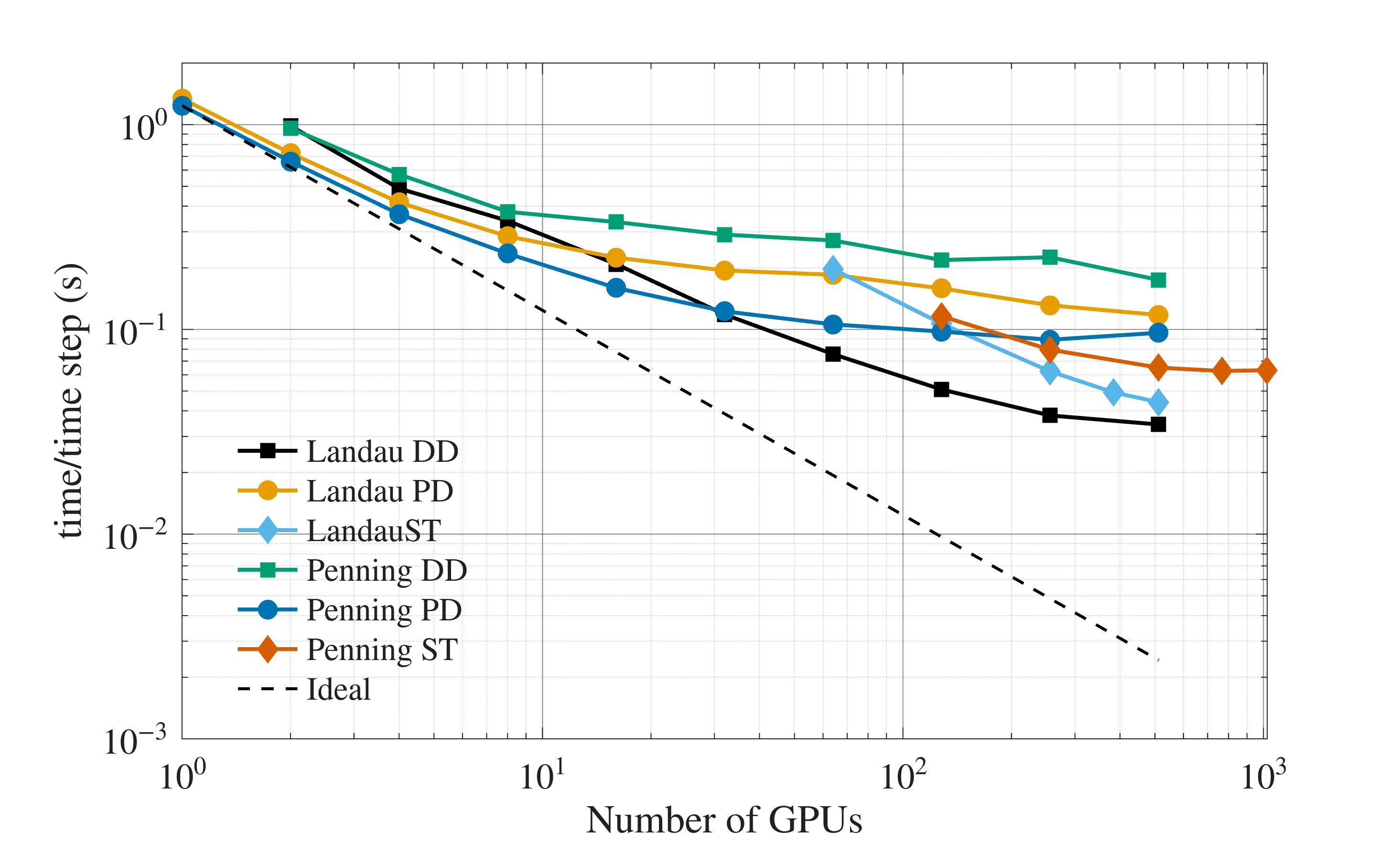}
\figlab{time_juwels}}
\subfigure[Speedup Alps]{\includegraphics[width=0.5\columnwidth]{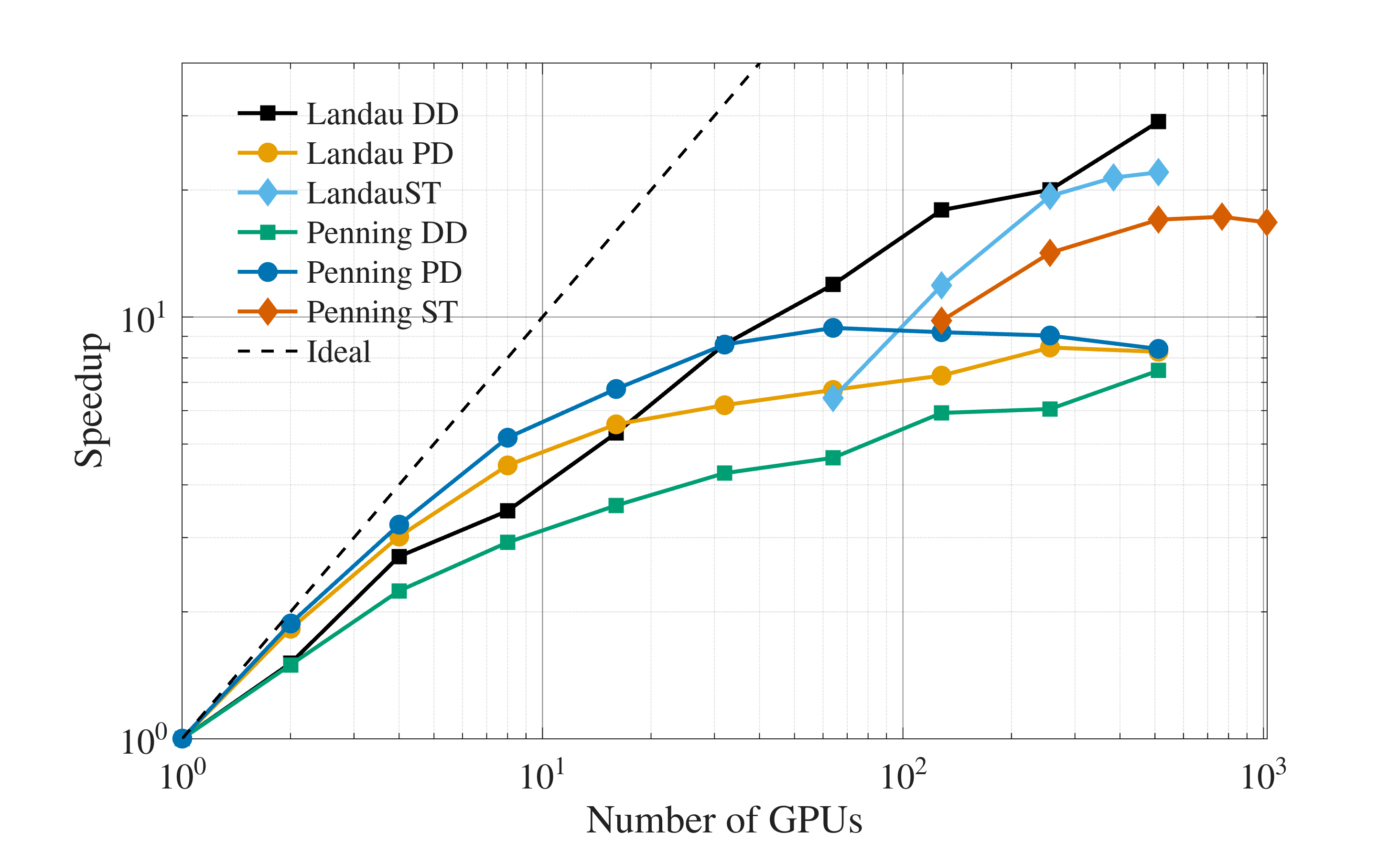}
\figlab{speedup_alps}}
\subfigure[Speedup JUWELS Booster]{\includegraphics[width=0.5\columnwidth]{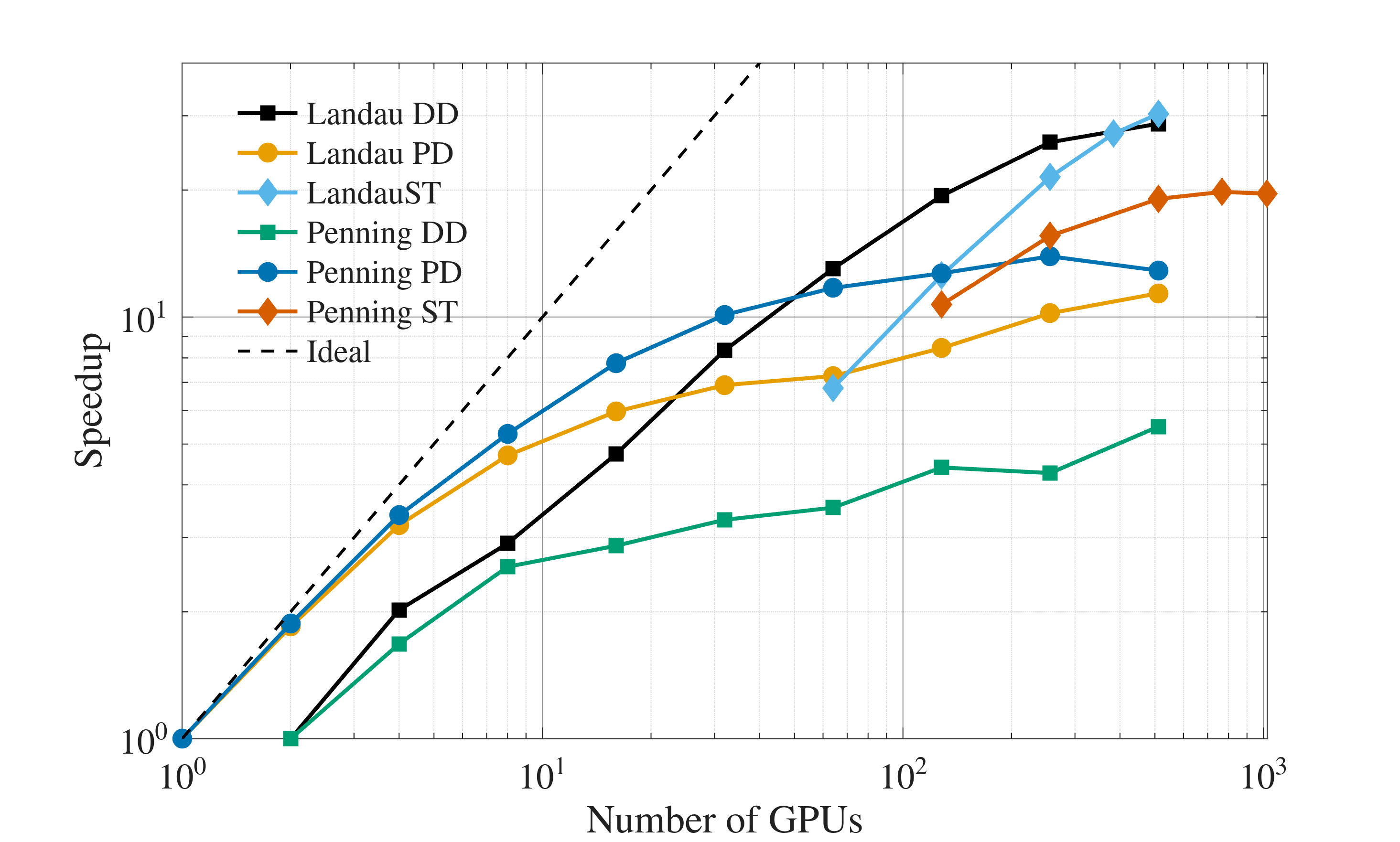}
\figlab{speedup_juwels}}
\subfigure[Efficiency Alps]{\includegraphics[width=0.5\columnwidth]{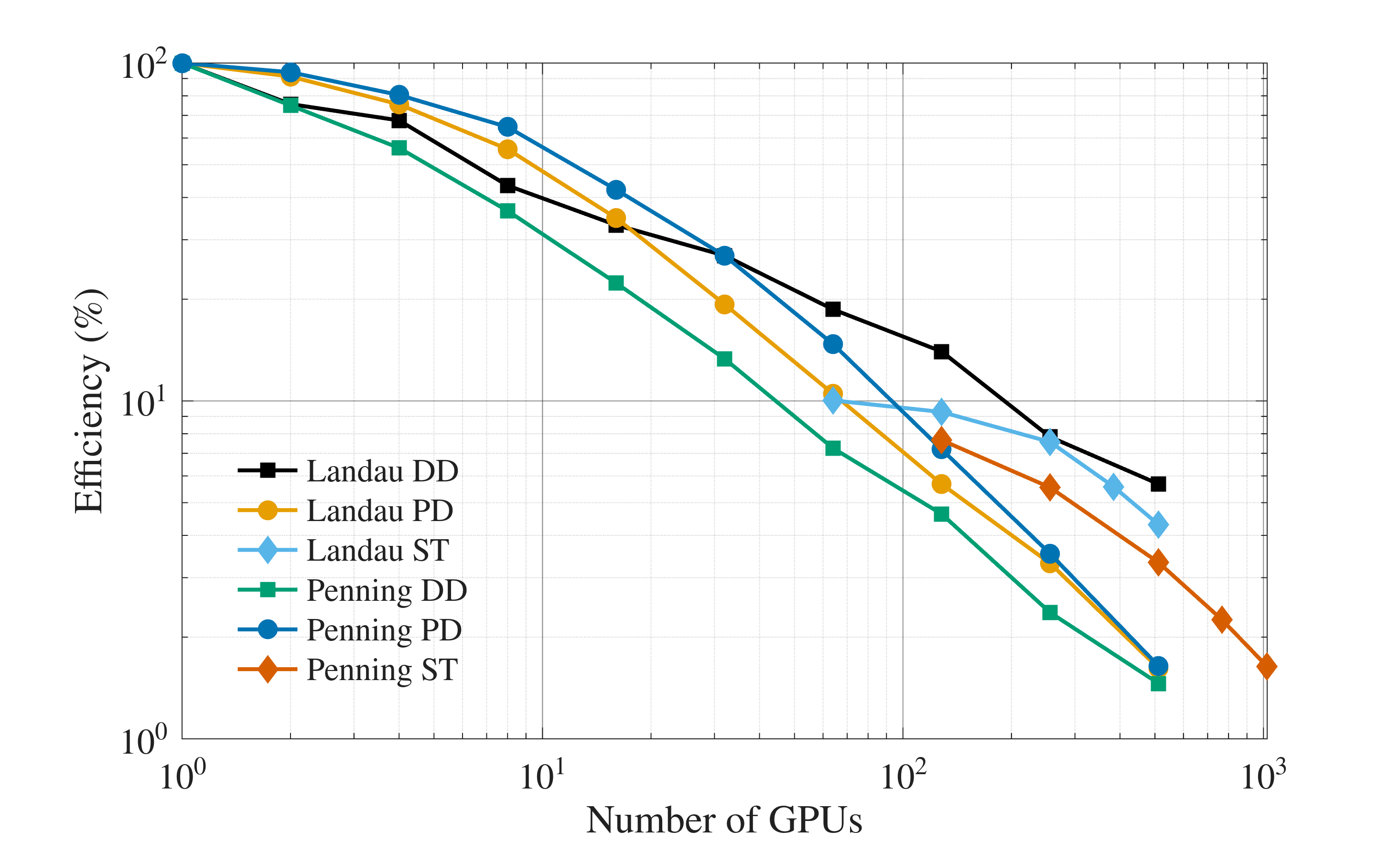}
\figlab{eff_alps}}
\subfigure[Efficiency JUWELS Booster]{\includegraphics[width=0.5\columnwidth]{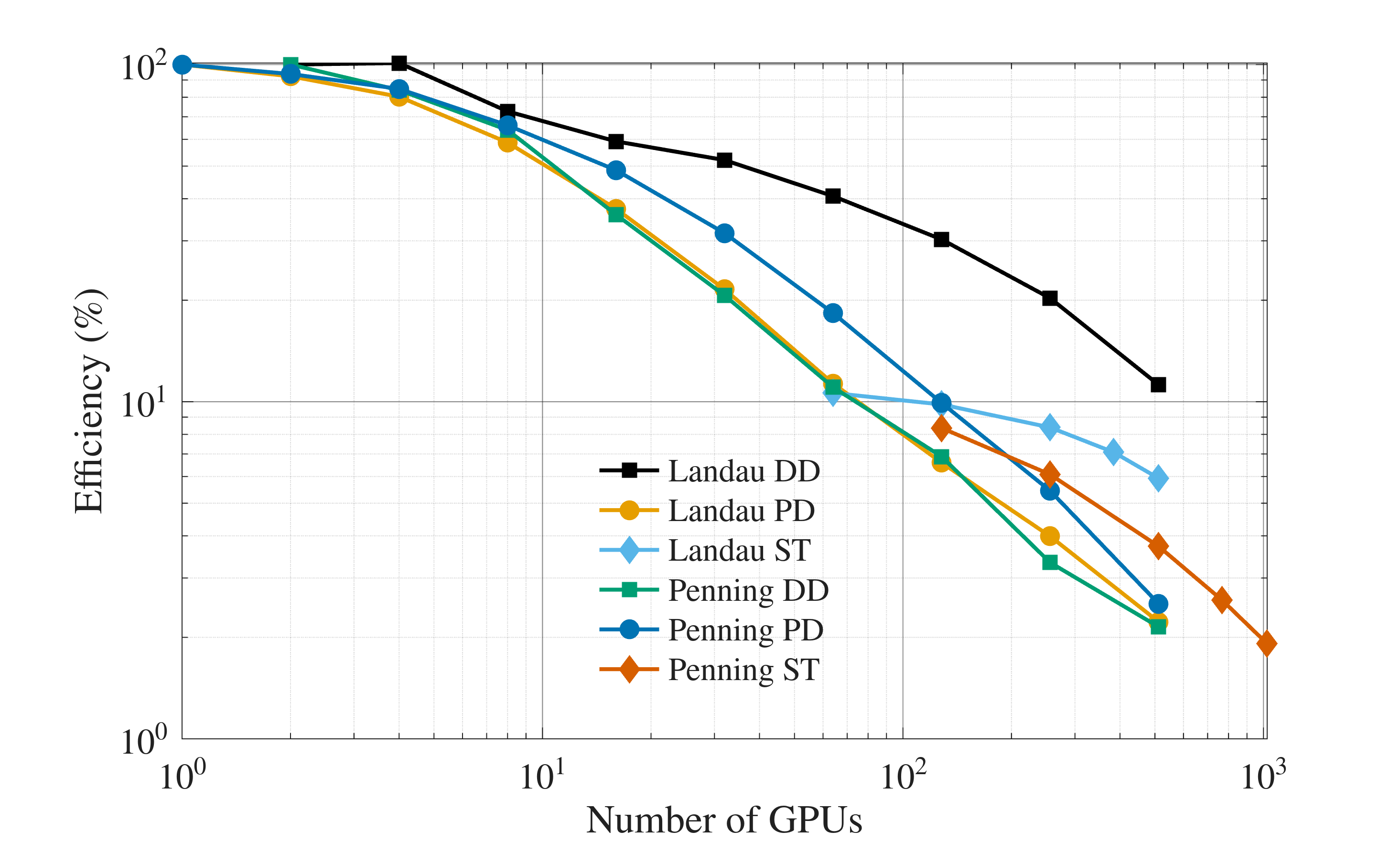}
\figlab{eff_juwels}}
\caption{Strong scaling study comparing different parallelization strategies for Landau damping and Penning trap test cases on Alps and JUWELS booster. The problem size is $256^3$ modes and 10 particles per mode. In the legends, DD stands for domain decomposition, PD for particle decomposition and ST for space-time decomposition. For the space-time decomposition we used the first data point of particle decomposition for the calculation of speedups and efficiencies, whereas for domain and particle decompositions we used the first available data points of the respective strategies.}
\figlab{Alps_juwels_overall}
\end{figure}

\subsubsection{Landau damping}

In Figure \figref{Alps_juwels_overall}, we show the strong scaling of the three strategies for the two test cases on Alps and JUWELS booster. On Alps, for Landau damping, domain decomposition is the best strategy especially at high number of GPUs. Particle decomposition performs slightly better compared to domain decomposition up to 8 GPUs, with the cross over point at 16 GPUs, after which it no longer scales well. Following \cite{muralikrishnan2025parapif}, we use the space-time decomposition once the parallel efficiency in spatial parallelization goes down roughly below fifty percent, which for Landau damping happens at 8 GPUs. 
For the coarse propagator in parareal, we use PIF with a tolerance of $\varepsilon=10^{-3}$ and a coarse time step size of $0.05$, which has been identified as the optimal combination in \cite{muralikrishnan2025parapif}. The parareal method takes at least 3-4 iterations to converge, hence it is not better than serial time stepping at 2 and 4 ranks for time parallelization. For this reason, the first data point for space–time decomposition is reported at 64 GPUs, with eight GPUs used for spatial parallelization (via particle decomposition) and eight GPUs used for time parallelization. At $8\times8$ GPUs, space-time decomposition performs similiarly to particle decomposition, and after that it scales and decreases the gap between particle decomposition and domain decomposition. At $8\times32$ GPUs, the runtime is nearly the same as that of domain decomposition for the same total number of GPUs. Beyond this point, scaling saturates due to communication overhead in the time dimension.

\subsubsection{Penning trap}

The results for the Penning trap benchmark, which has a clustered particle distribution, differ significantly from those for Landau damping. Here, up to 8 GPUs, the timings for domain decomposition are very similar to those for Landau damping, as the particles are load balanced. From 16 GPUs onward, load imbalance in particles affects the scaling. One could adopt particle load balancing strategies such as orthogonal recursive bisection to help alleviate this issue. However, as already noted in \cite{muralikrishnan2024scaling} for PIC schemes, this creates a load imbalance in fields, which then affects the scaling. For the NUFFTs in PIF, the timings are dominated by distributed uniform FFT timings at high core counts. Therefore, a load imbalance in fields would significantly affect the scaling, too. Load balancing strategies for PIF schemes which carefully take into account both particles and fields are still an active are of research and a subject of future work. 

Particle decomposition, on the other hand, is not affected by the nonhomogeneous particle distribution, thanks to its inherent load balancing. It even scales slightly better than in the Landau damping case. Apart from scaling, the absolute timings are also better due to FINUFFT handling clustered distributions better than homogeneous distributions. Hence, even at 8 GPUs, where the domain decomposition is still load balanced, the particle decomposition is approximately $2\times$ faster than domain decomposition. Thereafter the gap increases further until 64 GPUs as the load imbalance starts affecting domain decomposition. Beyond 64 GPUs, the scaling in particle decomposition also saturates from high communication and serial computation costs of fields and hence both the curves converge. 

Using a similar fifty percent criteria on parallel efficiency, we use $16$ GPUs for spatial parallelization in space-time decomposition for Penning trap. For the coarse propagator, we use the best combination from \cite{muralikrishnan2025parapif}, which is the standard PIC scheme with the same coarse time step as the fine time step size and the parareal method applied in 16 blocks over the whole time domain. This multi-block parareal method is needed for Penning trap because of the multiple time scales involved in it, which makes parareal effective only within a shorter time window. At $8$ GPUs for time parallelization ($16\times8$ total GPUs), similar to Landau damping, space-time decomposition has a runtime comparable to that of particle decomposition. Thereafter, the timings decrease, but the scaling is worse than that for Landau damping due to the increased communication costs involved in the multi-block parareal. Beyond 32 GPUs in the time dimension ($16\times32$ total GPUs), scaling saturates, similar to the behaviour observed for Landau damping. The space-time decomposition offers $2.5\times$ and $2\times$ speedups compared to domain decomposition and particle decomposition respectively at $512$ GPUs. Thus particle decomposition at low numbers of GPUs and space-time decomposition at high numbers of GPUs offers an alternative to domain decomposition and works robustly across homogeneous and nonhomogeneous particle distributions. 

On JUWELS booster in Figure \figref{time_juwels}, the inferences are mostly similar to Alps. The absolute runtimes are $1.5-2\times$ slower than those on Alps, which is expected given the performance charactetristics of NVIDIA GH200 systems compared to the A100. The domain decomposition scaling for Landau damping is better leading to greater differences between penning trap and Landau damping after 8 GPUs. This can be attributed to the Infiniband network on JUWELS booster compared to Slingshot on Alps. We ran out of memory on 1 GPU for the domain decomposition, whereas in particle decomposition we are still able to run. This is due to the slightly higher memory requirements of the native NUFFT implementation in IPPL than FINUFFT, which does not show up on Alps due to larger 96 GB memory available in GH200 compared to 40GB in A100. 

\subsection{Component timings}

\begin{figure}[h!b!t!]
\subfigure[Landau DD]{\includegraphics[width=0.5\columnwidth]{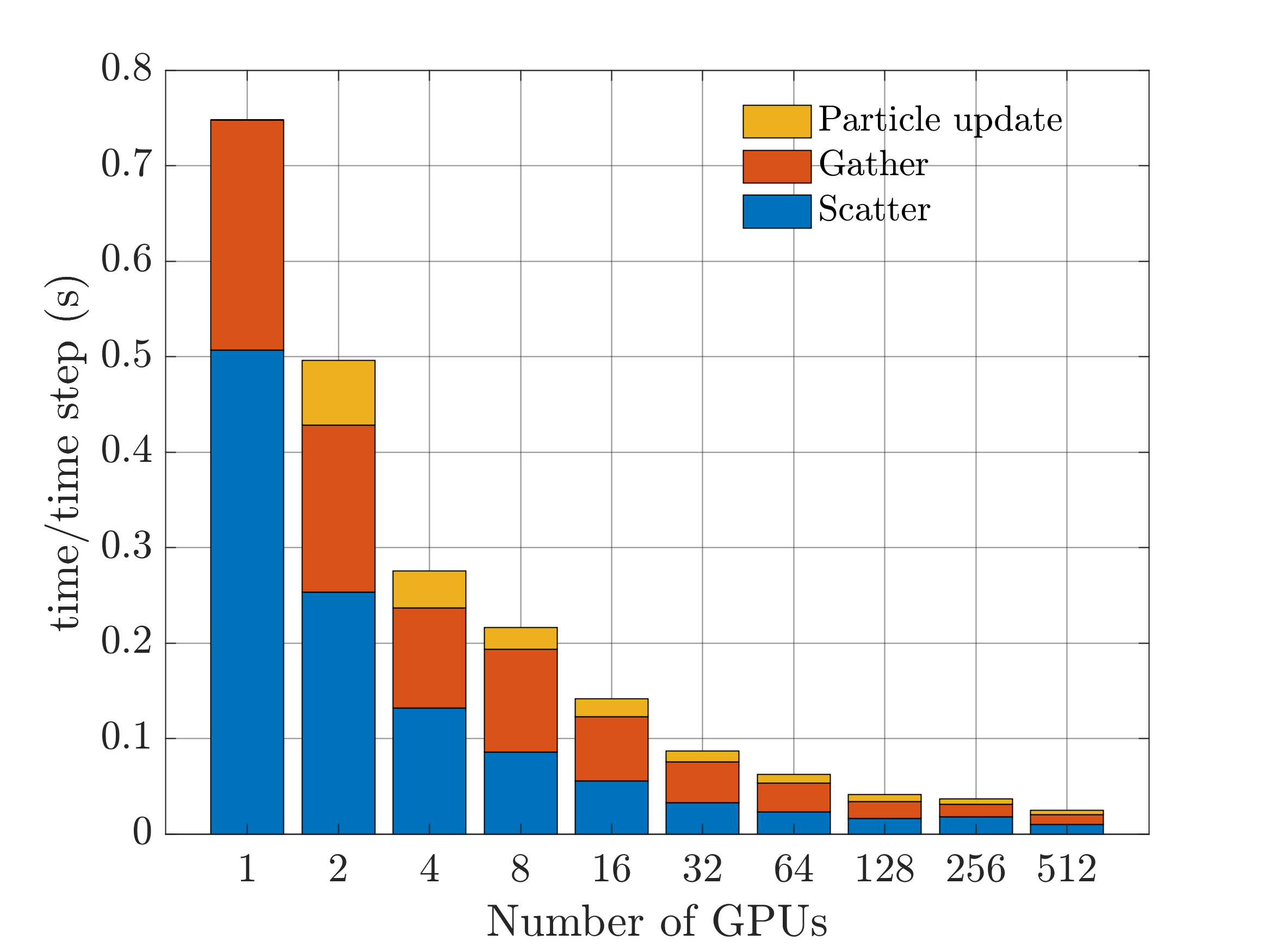}
\figlab{landau_dd_alps}}
\subfigure[Penning DD]{\includegraphics[width=0.5\columnwidth]{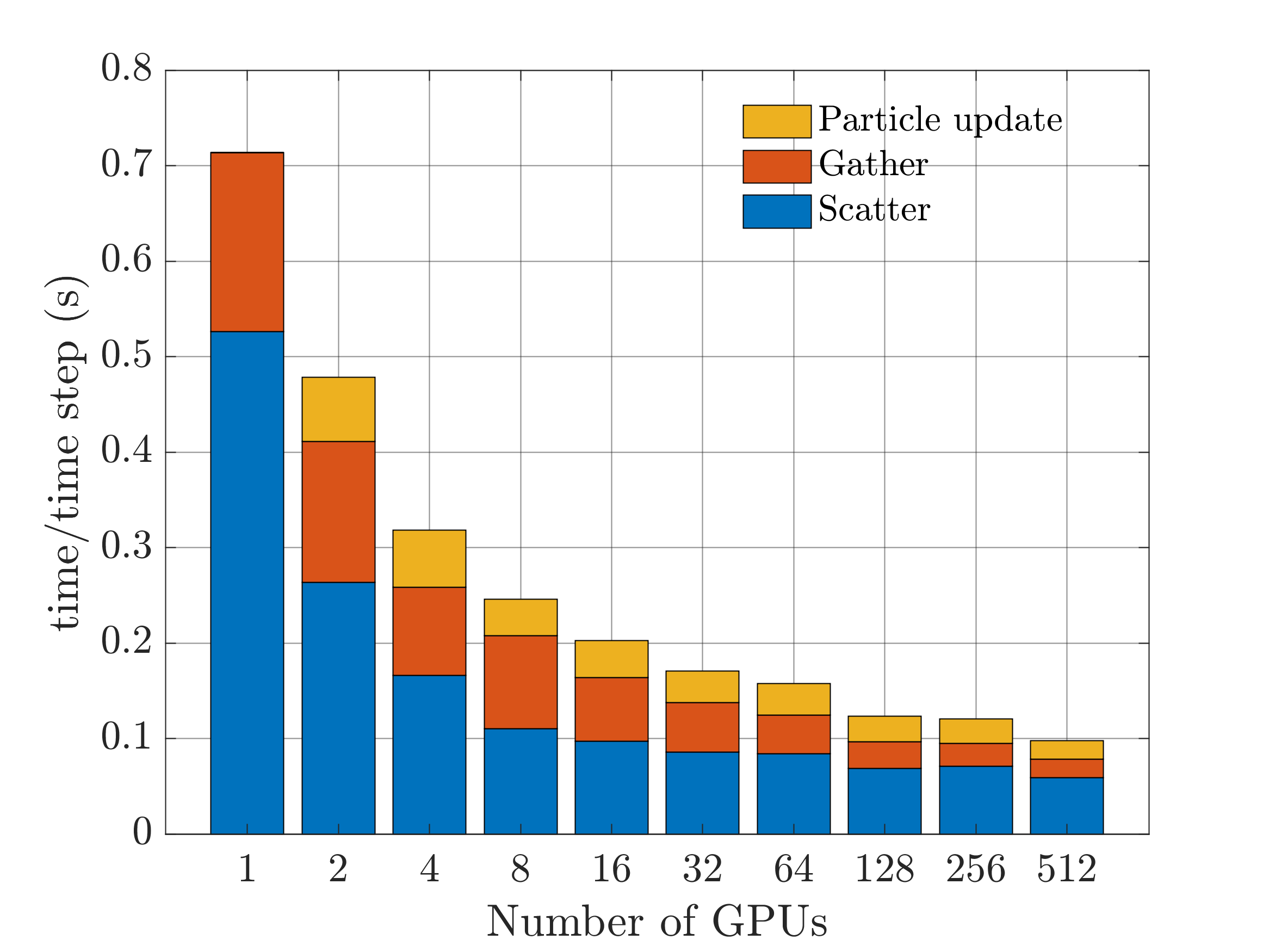}
\figlab{penning_dd_alps}}
\subfigure[Landau PD]{\includegraphics[width=0.5\columnwidth]{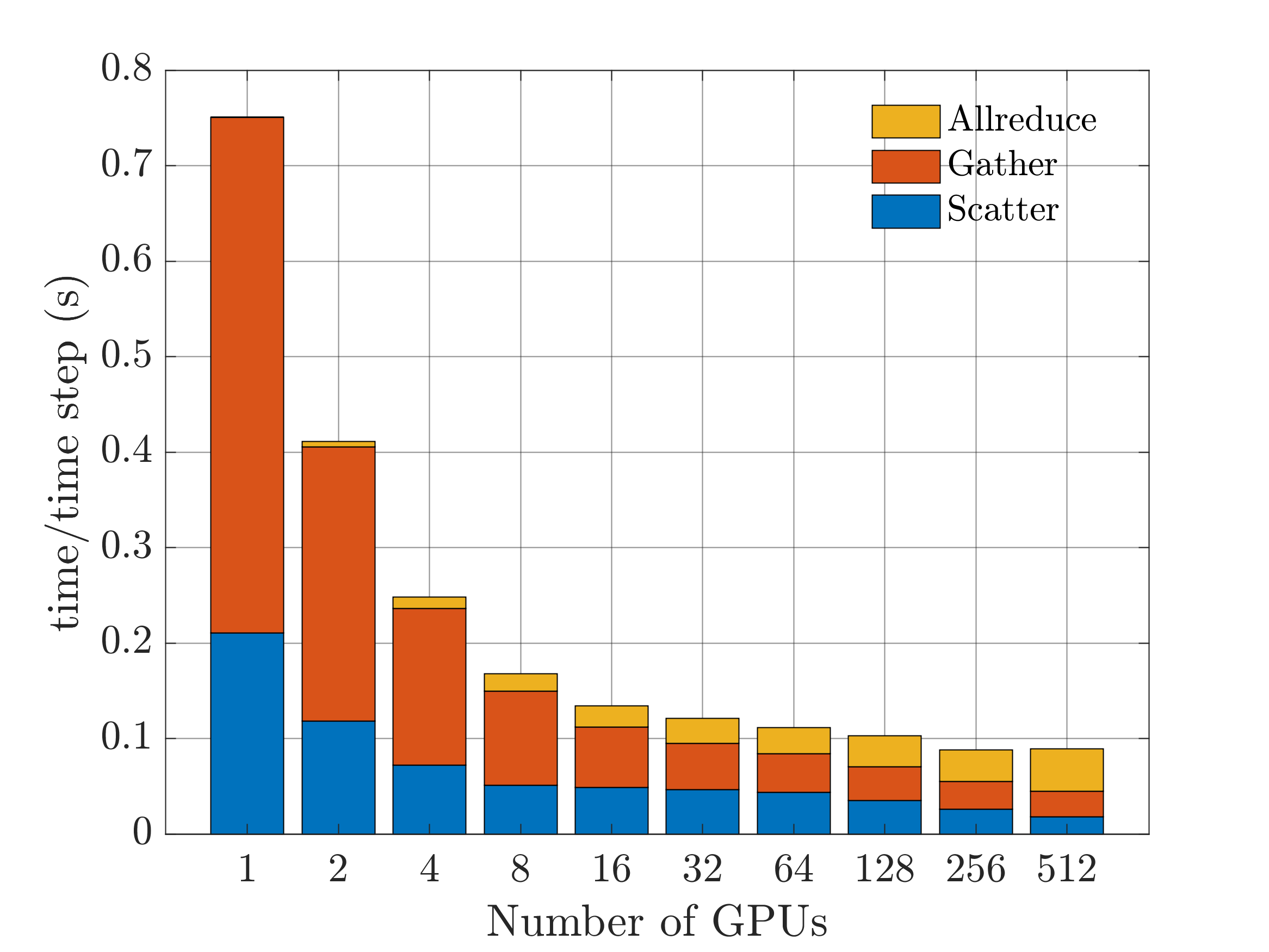}
\figlab{landau_pd_alps}}
\subfigure[Penning PD]{\includegraphics[width=0.5\columnwidth]{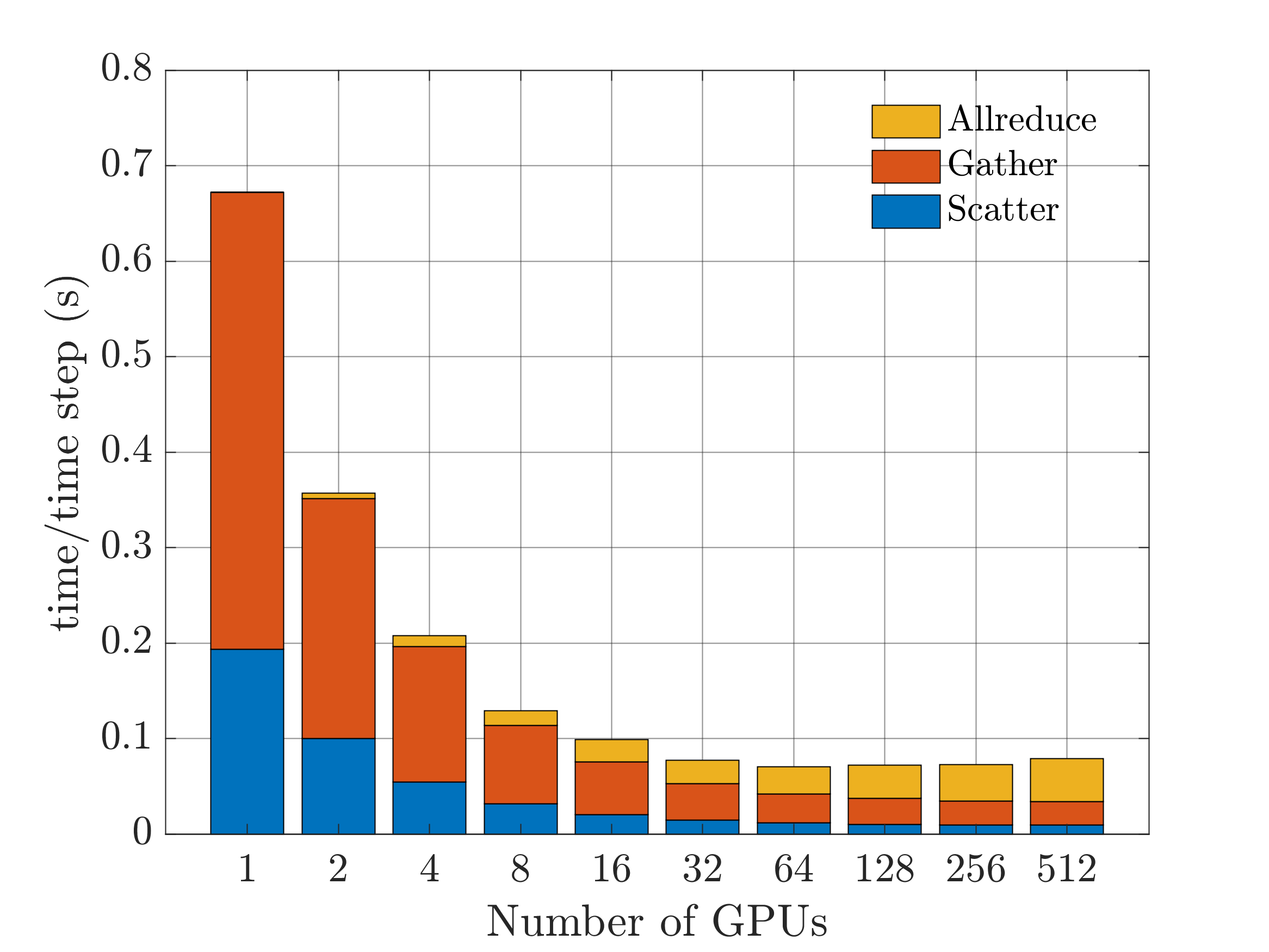}
\figlab{penning_pd_alps}}
\subfigure[Landau ST]{\includegraphics[width=0.5\columnwidth]{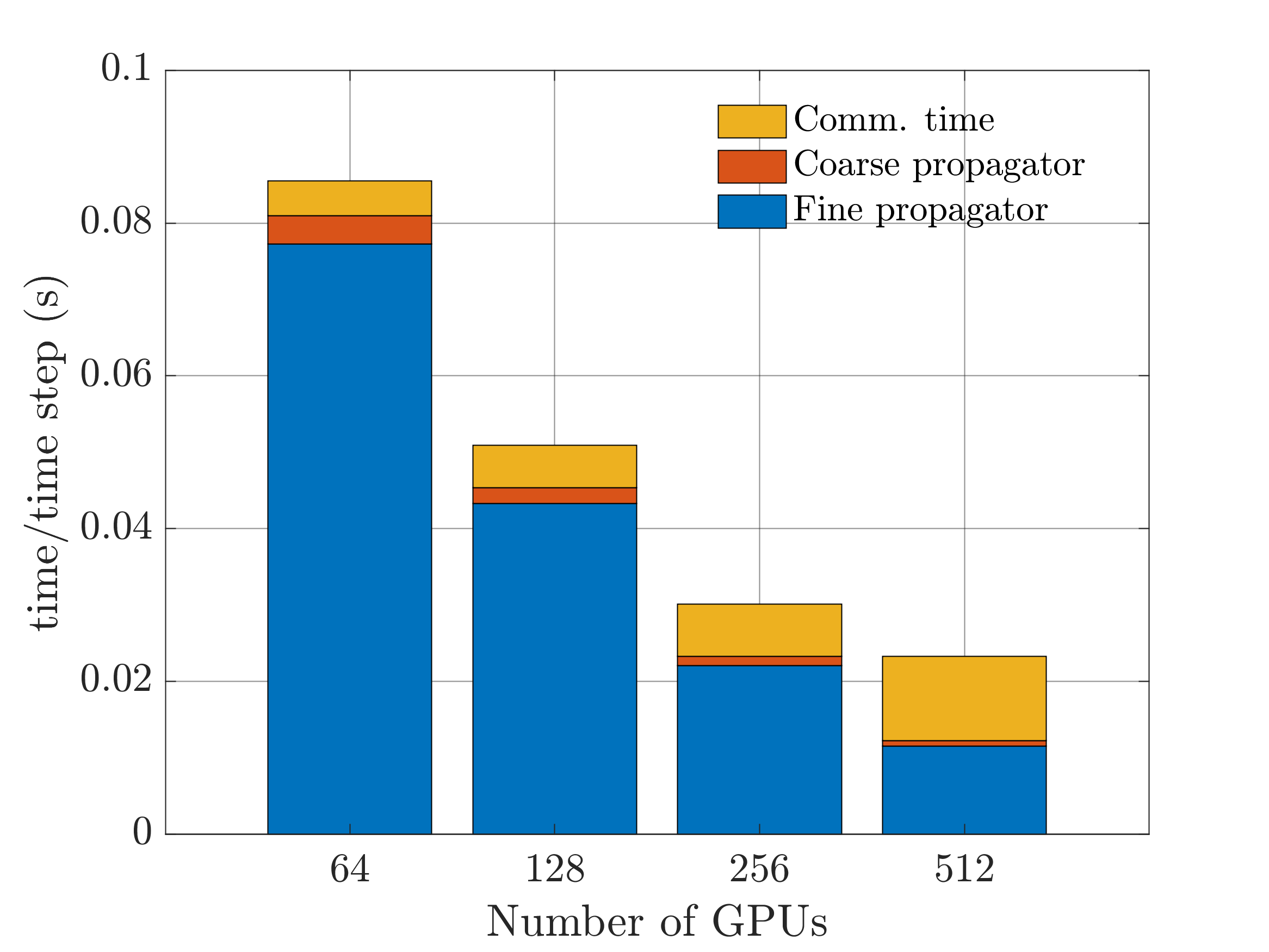}
\figlab{landau_st_alps}}
\subfigure[Penning ST]{\includegraphics[width=0.5\columnwidth]{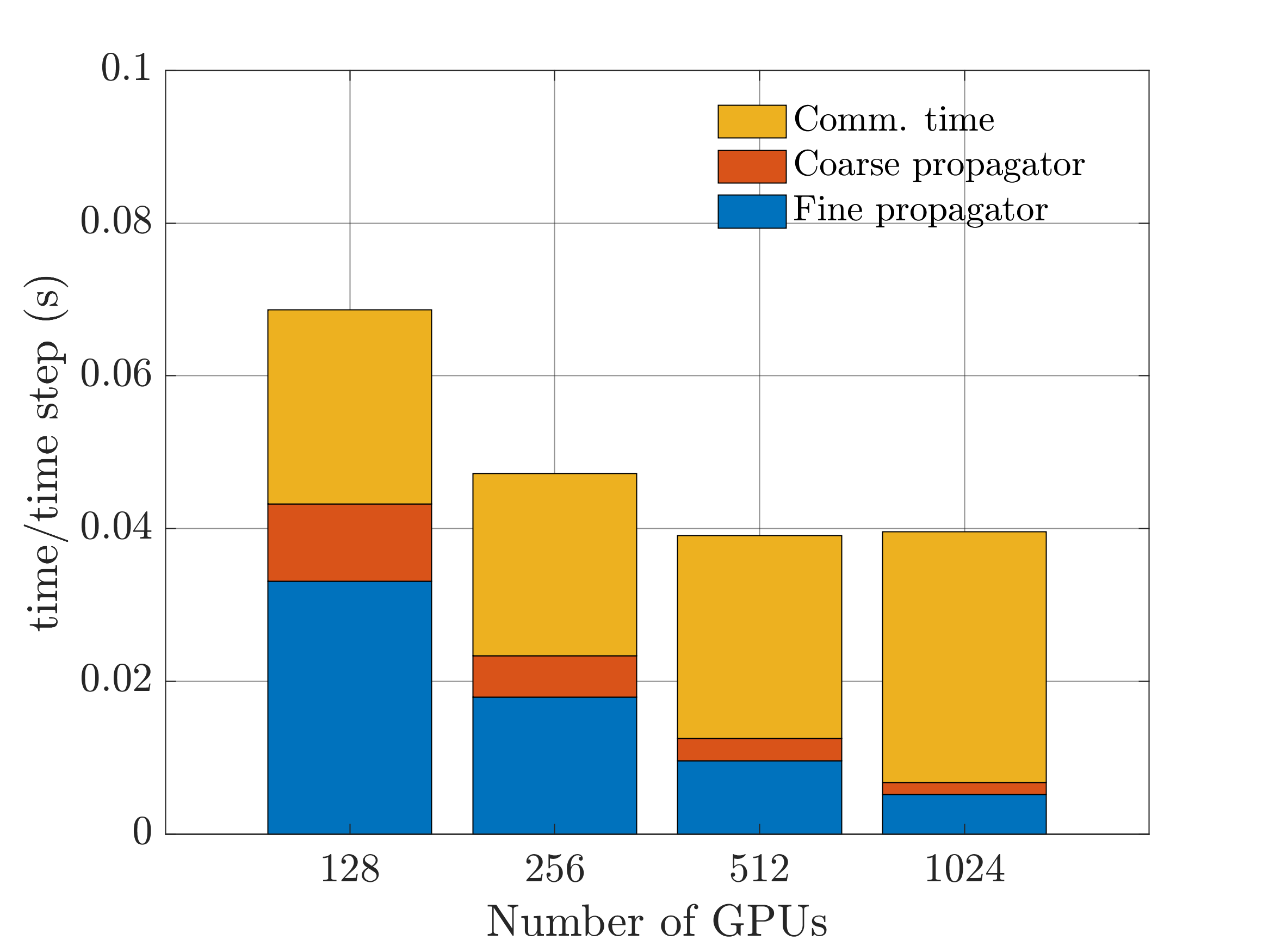}
\figlab{penning_st_alps}}
\caption{Strong scaling of dominant component timings of the three parallelization strategies for Landau damping and Penning trap test cases on Alps. In the sub-figure labels, DD stands for domain decomposition, PD for particle decomposition and ST for space-time decomposition.}
\figlab{Alps_comp}
\end{figure}

\subsubsection{Dominant kernels}

In this section, we analyze the scaling and performance of the three dominant components in each of the parallelization strategy that account for more than ninety percent of the total timings. The ``Scatter'' and ``Gather'' operations are two of the dominant components in both domain decomposition and particle decomposition. The Poisson solve is embedded into the gather routine itself in order to save memory by not storing the electric field in Fourier space. This is important especially for particle and space-time decompositions, as the fields are not distributed in those cases. Scatter and gather operations involve type 1 and type 2 NUFFTs, the complexities of both being $\mc{O}\LRp{(|log \varepsilon|+1)^d N_p + N_m log N_m}$. The scatter step performs a single type 1 NUFFT, whereas the gather step performs three type-2 NUFFTs---one for each component of the electric field in three dimensions. For a total number of $P$ MPI ranks, in domain decomposition each rank carries $N_p/P$ and $N_m/P$ local particles and Fourier modes, whereas in particle decomposition each rank has $N_p/P$ local particles and all the $N_m$ Fourier modes. The timings of scatter and gather in domain decomposition include field halo communication as well as distributed FFT communication, whereas in particle decomposition they do not include any communication timings.  The third component in domain decomposition is ``Particle update'' which communicates particles to neighboring ranks once they leave their local spatial field domains. For the particle decomposition, it is the ``Allreduce'' communication of the Fourier modes that collects the local scatter contributions from each rank.

The dominant components in the space-time decomposition are i) ``Fine propagator'' which includes all the timings in the fine level PIF time stepping, ii) ``Coarse propagator'' which includes all the timings in the coarse level PIF or PIC time stepping and iii) ``Communication time'' which accounts for particle communication time in the time communicator.

\subsubsection{Domain decomposition}

Figure \figref{Alps_comp} shows the component timings on Alps. We plot the average runtime of the different components across the MPI ranks. For Landau damping with domain decomposition in Figure \figref{landau_dd_alps}, all the components exhibit good scaling. The most dominant component is scatter, which is followed by gather and then the particle update. For the Penning trap benchmark in Figure \figref{penning_dd_alps}, the component timings look similar to those of Landau damping up to 8 GPUs and, after that, the scaling deteriorates due to particle load imbalance. 

\subsubsection{Particle decomposition}

For the particle decomposition in Figures \figref{landau_pd_alps} and \figref{penning_pd_alps}, gather is the dominant component, followed by scatter and then the allreduce. It is interesting to note that even though the total timings on 1 GPU remain similar between domain decomposition and particle decomposition, in the domain decomposition, which uses the native NUFFT in IPPL, scatter is the dominant component, whereas for the particle decomposition, which uses FINUFFT, the ratio is reversed. This is due to the differences in the algorithmic choices and implementation details between FINUFFT and IPPL's native NUFFT. The scaling deteriorates after 8 GPUs in particle decomposition as it parallelizes only over particles and we have 10 particles per mode. Hence, from 16 GPUs onward the computational costs associated with the serial fields dominate and inhibit the scaling. The cost of allreduce operations associated with the fields grow with number of ranks, as expected. The scaling behaviour and absolute component timings in the Penning trap case in Figure \figref{penning_pd_alps} are slightly better than those of Landau damping in \figref{landau_pd_alps}, due to the differences in particle distributions as described before. 

\subsubsection{Space-time decomposition}

In the case of space-time decomposition for Landau damping in Figure \figref{landau_st_alps}, the dominant component is the fine propagator. Both the fine and coarse propagator timings decrease with the number of ranks, whereas the communication time grows. There is a fundamental difference between the particle communication timings in domain decomposition in Figure \figref{landau_dd_alps} and space-time decomposition in Figure \figref{landau_st_alps}. In the former, the number of spatial neighbors remains constant and does not increase with the number of ranks. Since the local number of particles decreases in domain decomposition, the number of particles being communicated with the neighbors also decreases, and hence the communication time decreases. On the other hand, in time parallelization, the communication occurs sequentially over the time ranks owing to the causality of time. Hence, the runtime increases with a higher number of ranks, as the local number of particles remains the same. For the Penning trap in Figure \figref{penning_st_alps}, the fine and coarse propagator timings scale in the same way as in Landau damping. The absolute timings are lower due to the clustered distribution. However, the communication time is much longer and dominates the overall timings starting from 256 GPUs. This is due to the multi-block parareal algorithm, which parallelizes only over a small time window and the blocks then proceed sequentially in time once parareal converges within a block.

The component timings on JUWELS booster are shown in Appendix \secref{appendix} and the inferences are similar to that of  Alps except for the absolute timings being $1.5-2\times$ higher. 

\section{Conclusion}
\seclab{conclusion}

We present and analyze three distributed parallelization strategies for particle-in-Fourier (PIF) schemes applied to the Vlasov-Poisson system in kinetic plasma simulations. 

The first strategy is domain decomposition, where both particles and Fourier modes are distributed across MPI ranks. It is both memory- and work-optimal, and suited for applications which need both large numbers of particles as well as Fourier modes. It involves a dynamic communication pattern and needs efficient buffer management for scalability and performance on GPUs. It also requires combined particle and field load balancing strategies for handling nonhomogeneous particle distributions. It is the most intrusive strategy among the three of them, but can reuse machinery of standard particle-mesh codes for implementation.  

The second strategy is particle decomposition, where only the particles are distributed while the Fourier modes are duplicated in every rank. It is neither memory- nor work-optimal and suited for applications with high particle densities and a relatively low number of Fourier modes which can fit on a single GPU. It involves a simple, static communication pattern and no buffer management. It is field and particle load balanced by design and hence relatively agnostic to particle distributions. It is the least intrusive among the strategies and can make use of available open-source shared memory NUFFT libraries for implementation.

The third strategy is space-time decomposition, where parareal-based time parallelization is added on top of particle decomposition. It helps to improve the scalability and time-to-solution beyond particle decomposition by parallelizing the work in the time direction. It is also not memory- and work-optimal and involves the most memory and redundant computations in space and time among the three strategies. The communication pattern is static and the buffers can be pre-allocated once and used for the rest of the simulation. It is also particle and field load balanced by design, but its effectiveness depends on the spatio-temporal dynamics of the problem. In terms of intrusiveness, it stands between domain decomposition and particle decomposition. 

We implement the strategies in the open-source, C++ library IPPL and perform a strong scaling study with two benchmark problems, Landau damping and Penning trap, on the Alps and JUWELS booster supercomputers. As expected, domain decomposition works best for Landau damping which has a homogeneous particle distribution whereas for Penning trap, which has a clustered distribution, it suffers from load imbalance. Particle decomposition at low numbers of GPUs and
space--time decomposition at high numbers of GPUs offer competitive time-to-solution compared
to domain decomposition for Landau damping. For Penning trap, we obtain up to $2.5\times$ faster
time-to-solution compared to domain decomposition.

In terms of future work, particle and field load balancing strategies need to be studied for domain decomposition. Point-to-point communication involved in the particle migration of domain decomposition can be overlapped with the scatter operation using higher order kernels in type-1 NUFFT allowing the possibility to overlap communication with computation. This will be explored in a future work. Time parallelization currently suffers from high communication times for test cases like Penning trap which has multiple time scales. In an ongoing work, we are investigating the application of multigrid reduction in time~ \cite{falgout2014parallel}, which is a multilevel time parallelization strategy, towards alleviating this issue.

\backmatter


\bmhead{Acknowledgements}

We acknowledge the computing time on JUWELS booster through the project CSTMA and on Alps through the PASC project c41.

\bmhead{Availability}
IPPL is an open source project.\ The source code can be found here: 
\href{https://github.com/IPPL-framework/ippl}{https://github.com/IPPL-framework/ippl}. The branch used for this study can be found in the following tag: \href{https://github.com/srikrrish/ippl/releases/tag/go20-R0}{https://github.com/srikrrish/ippl/releases/tag/go20-R0}. 
On Alps, we use the following user environment: \texttt{/capstor/store/cscs/cscs/public/uenvs/\\ 
opal-x-gh200-mpich-gcc-2025-09-28-cuda12.8.squashfs}. On JUWELS Booster, we load the following modules: \texttt{Stages/2025 GCC OpenMPI CMake NCCL}. We use \texttt{Kokkos} version 5.0.0 and \texttt{heFFTe} version 2.4.1.

\begin{appendices}

\section{Component timings on JUWELS booster}
\seclab{appendix}

Here, we show the component timings for the three strategies on JUWELS booster.

\begin{figure}[h!b!t!]
\subfigure[Landau DD]{\includegraphics[width=0.5\columnwidth]{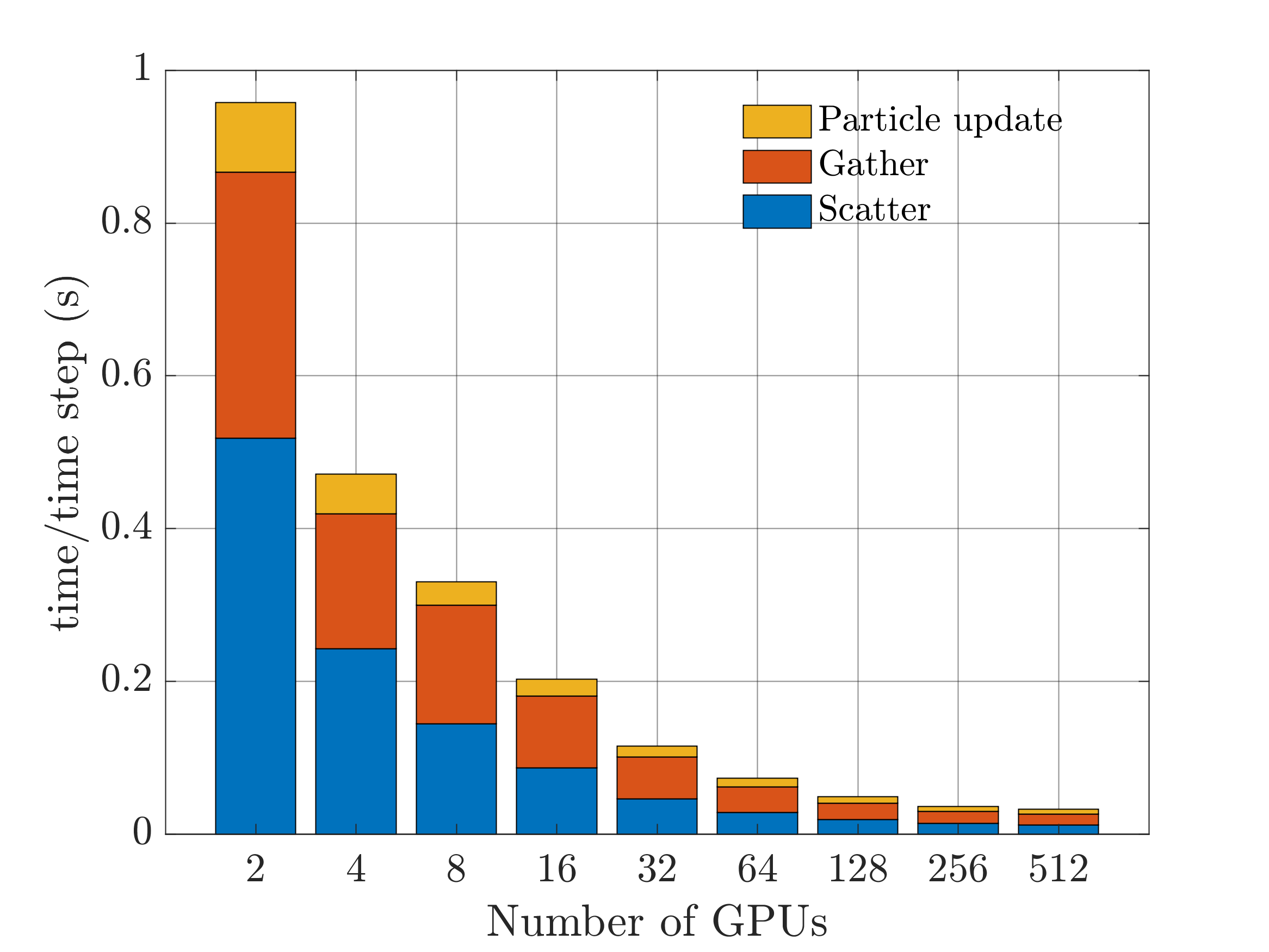}}
\subfigure[Penning DD]{\includegraphics[width=0.5\columnwidth]{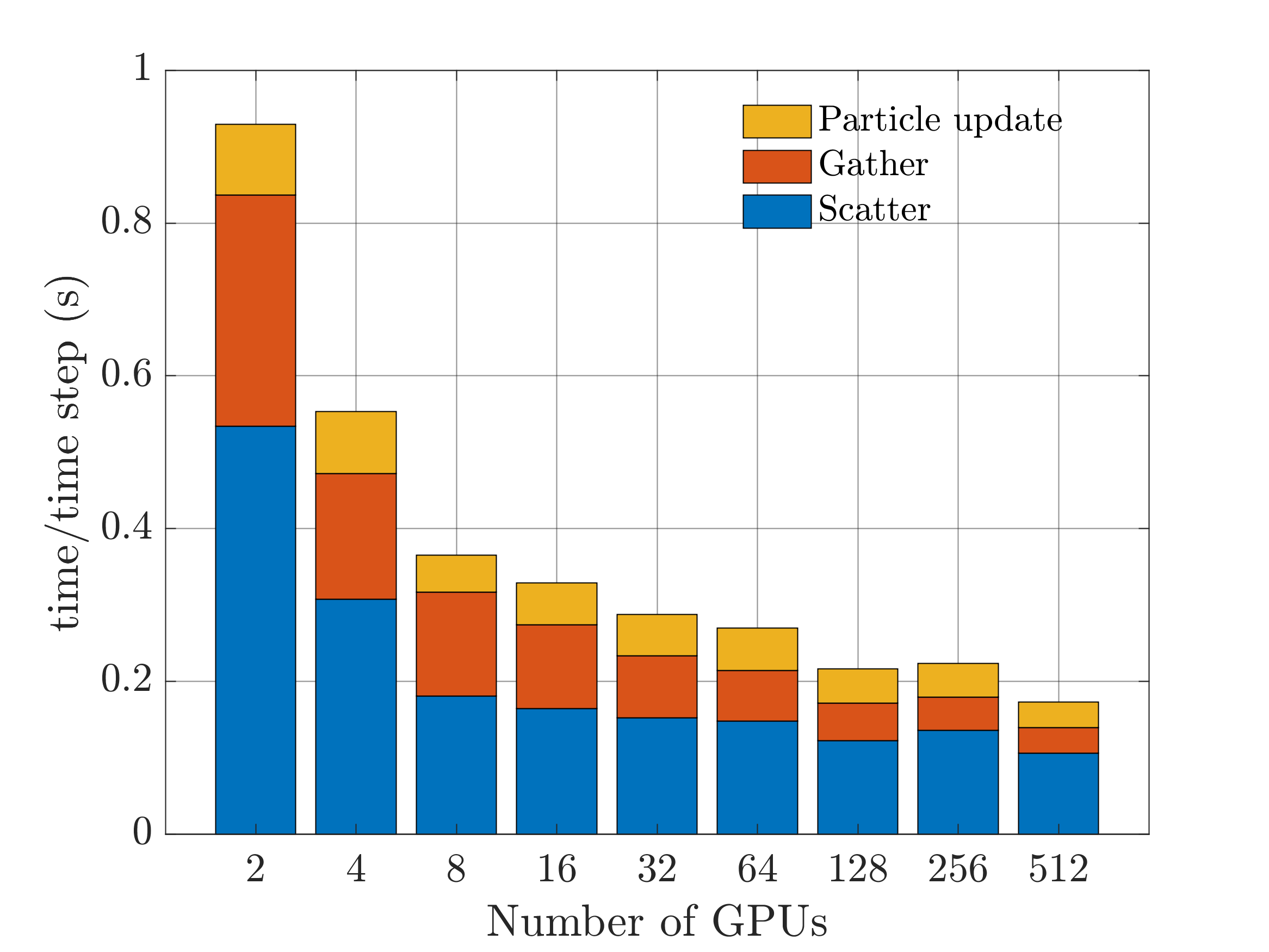}}
\subfigure[Landau PD]{\includegraphics[width=0.5\columnwidth]{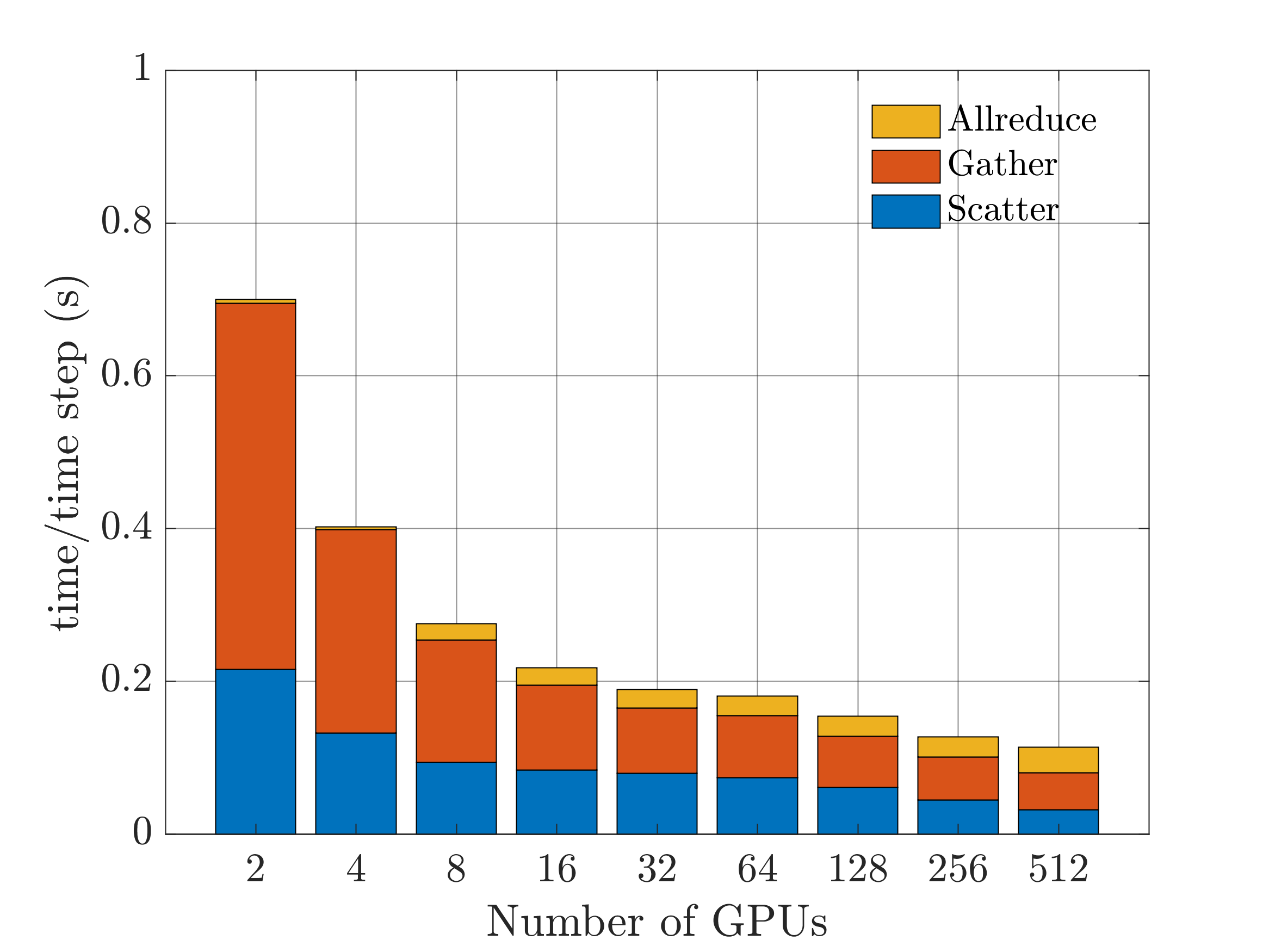}}
\subfigure[Penning PD]{\includegraphics[width=0.5\columnwidth]{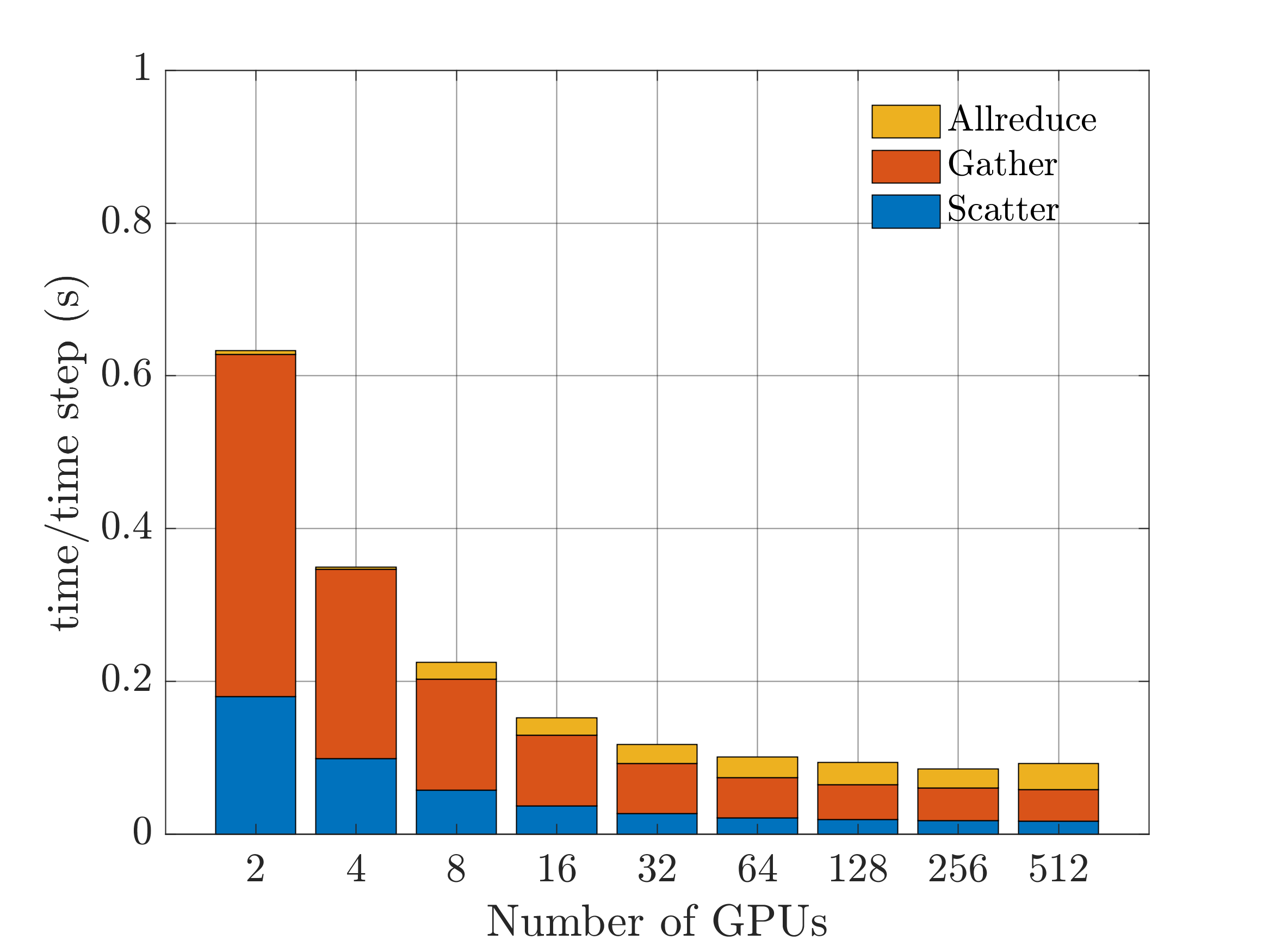}}
\subfigure[Landau ST]{\includegraphics[width=0.5\columnwidth]{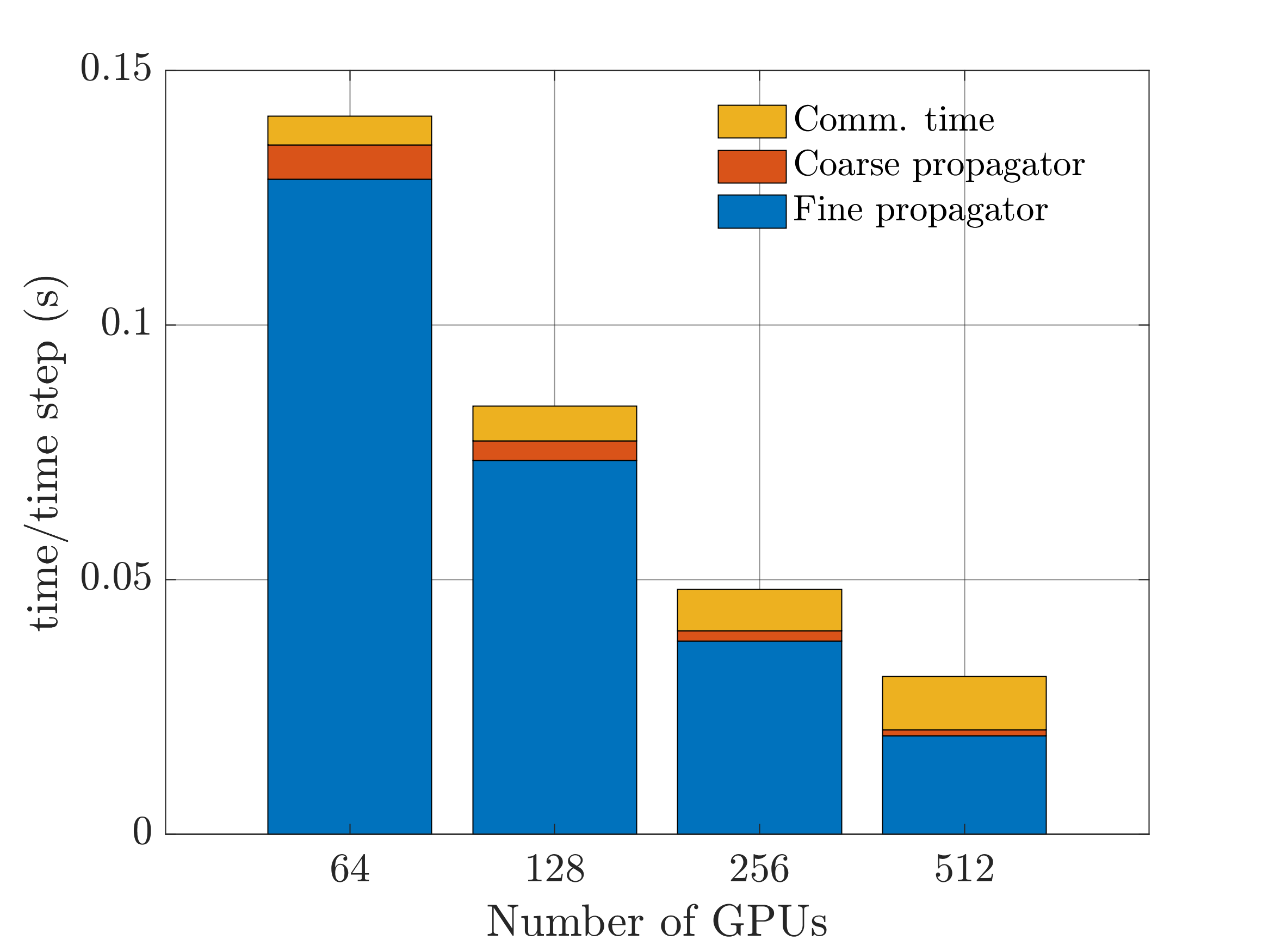}}
\subfigure[Penning ST]{\includegraphics[width=0.5\columnwidth]{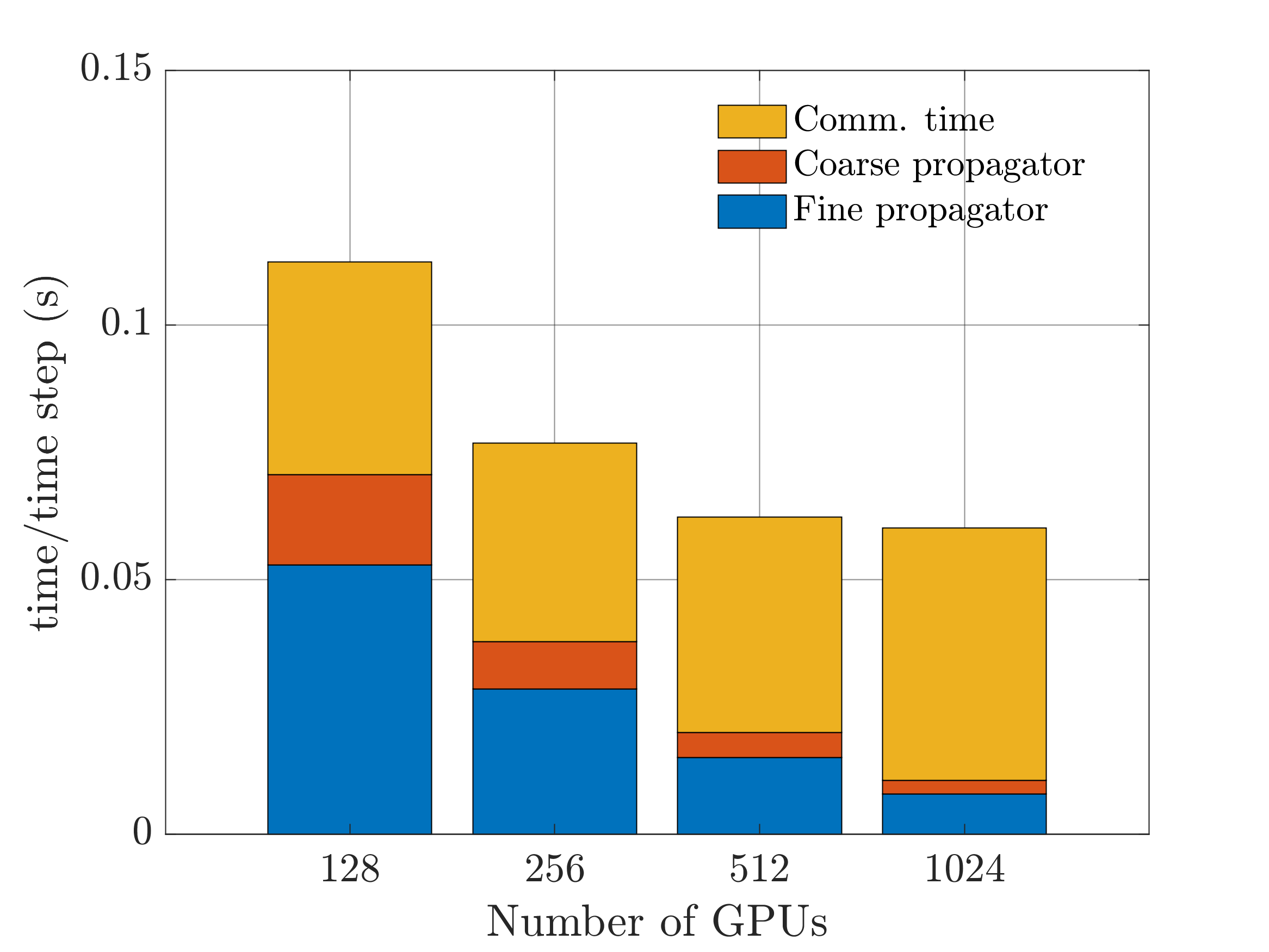}}
\caption{Strong scaling of dominant component timings of the three parallelization strategies for Landau damping and Penning trap test cases on JUWELS booster. In the sub-figure labels, DD stands for domain decomposition, PD for particle decomposition and ST for space-time decomposition.}
\figlab{Juwels_comp}
\end{figure}

\end{appendices}

\bibliography{references,self}

\end{document}